\documentclass[preprint,12pt]{elsarticle}

\usepackage{amssymb}
\usepackage{amsmath}
\journal{Nuclear Physics A}

\newcommand{\be}{\begin{equation}}
\newcommand{\ee}{  \end{equation}}
\newcommand{\ba}{\begin{eqnarray}}
\newcommand{\ea}{  \end{eqnarray}}
\newcommand{\ve}{\varepsilon}

\begin{document}

\begin{frontmatter}

\title{Nuclear Level Densities at High Excitation Energies and for
Large Particle Numbers}

\author{Adriana P\'alffy\corref{cor1}}
\ead{palffy@mpi-hd.mpg.de}

\author{Hans A. Weidenm\"uller}
\ead{haw@mpi-hd.mpg.de}

\address{Max-Planck-Institut f\"ur Kernphysik, Saupfercheckweg 1, D-69117 Heidelberg, Germany}
\cortext[cor1]{Corresponding author. Tel.: +49(0)6221 516171, fax: +49(0)6221 516152}

\begin{abstract}
Starting from an independent-particle model with a finite and
arbitrary set of single-particle energies, we develop an analytical
approximation to the many-body level density $\rho_A(E)$ and to
particle-hole densities. We use exact expressions for the low-order
moments and cumulants to derive approximate expressions for the
coefficients of an expansion of these densities in terms of orthogonal
polynomials. The approach is asymptotically (mass number $A \gg 1$)
convergent and, for large $A$, covers about 20 orders of magnitude
near the maximum of $\rho_A(E)$ (i.e., about half the spectrum).
Densities of accessible states are calculated using the Fermi-gas
model.

\end{abstract}


\end{frontmatter}


\section{Motivation}

The theoretical description of nuclear reactions at high energy
typically uses rate equations. A prime example is provided by the
treatment of precompound or pre-equilibrium reactions, see
Refs.~\cite{Fes80, Bla85}. The rates involve the product of a basic
strength factor with either the density of accessible states, or with
the ratio of the total level density at different energies, or with the
ratio of level densities for different particle-hole numbers at the
same energy. In some cases of practical interest, such level densities
are needed at high excitation energies and for large particle
numbers. This is the case, for instance, for reactions between heavy
ions at energies of several MeV per nucleon~\cite{Mus93}. It is also
true for nuclear reactions induced by zeptosecond multi-MeV laser
pulses~\cite{Pal13}. The ``Nuclear Physics Pillar'' of the ``Extreme
Light Infrastructure'' (ELI)  \cite{Eli12,DiP12} holds promise to deliver
such pulses with coherent photons in the not--too--distant
future~\cite{Mou11}. These would excite target nuclei up to several
100 MeV above yrast. Beyond the dependence of the total level density
on excitation energy, spin and parity, one needs
densities of states with fixed particle-hole number or with fixed
numbers of protons and neutrons, densities of states accessible from a
particular particle-hole state, and similar related quantities.
The challenge here for already established methods 
is related to the high excitation energies  and the 
large particle-hole numbers involved.

In this paper we present a new method to calculate all these
quantities in a manner which is both transparent and easy to
implement. We use a single-particle model with non-interacting
Fermions containing a finite number of bound single-particle
states. These have definite quantum numbers (spin, angular momentum,
parity) and may either be taken from the empirical spherical shell
model or from a self-consistent calculation. We can also accommodate
results of a temperature-dependent Hartree-Fock calculation.

Generalizing the approach developed in Ref.~\cite{Pal13}, we start
from an exact expression for the total level density $\rho_A(E)$ in
terms of Fermionic occupation numbers. Here $E$ and $A$ stand for the
energy and particle number, respectively. A significant extension of
Ref.~\cite{Pal13} and a decisive step in the calculation is that we
take the Fourier transform of $\rho_A(E)$ with respect to energy $E$
and the Laplace transform with respect to particle number $A$. The
resulting closed-form expression for the double transform $L$ yields
exact expressions for the low moments and low cumulants of
$\rho_A(E)$. We approximate the Fourier transform of $\rho_A(E)$ in
terms of these cumulants and use the result to determine the
coefficients of an expansion of $\rho_A(E)$ in terms of orthogonal
polynomials. All these steps are carried through analytically. The
same steps are used to work out particle-hole densities for fixed
particle-hole numbers. The resulting expressions all depend on the
moments of the distribution of the single-particle energies. These
can be worked out analytically for models with continuous
single-particle level densities. Three such models are used to
demonstrate the success and the limitations of our approach: the
constant-spacing model used in Ref.~\cite{Pal13}, a model with linear
and a model with quadratic energy dependence of the single-particle
level density, suitable for a description of medium-weight and heavy
nuclei, respectively. In all three cases we can prove the rapid
convergence of our approximation scheme provided that $A \gg 1$. We
complete our approach by calculating densities of accessible states in
the framework of a modified Fermi-gas model that takes account of the
finiteness of the single-particle potential.

From the early days of nuclear physics, the nuclear level density has
attracted strong interest. Calculations of the nuclear level density
date back to the seminal work of Bethe~\cite{Bet36}. Particle-hole
densities were first worked out by Ericson~\cite{Eri60}. The very
substantial body of work that followed can roughly be grouped as
follows~\cite{Jac86}: (i) Exact combinatorial counting. Examples are
Refs.~\cite{Hil69, Ber74, Gho83, Bli86, Gri90, Hil98}. The approach
becomes intractable for large excitation energies and/or large
particle numbers. (ii) Gram-Charlier expansion around a
Gaussian-shaped level density. Examples are Refs.~\cite{Cha71,
  Das75}. The method fails badly in the vicinity of the ground-state
energy. (iii) Saddle-point approximations for the grand partition
function. Examples are Refs.~\cite{Blo68, Wil71, Sta85, Obl86,
  Her92}. To the best of our knowledge, this method has not been
applied to or tested for the large particle numbers and high
excitation energies considered in this paper. (iv) Recursive methods.
Examples are Refs.~\cite{Wil68, Alb73, Jac86}. These are exact and
work also for large particle numbers but provide numerical results
only. (v) Thermodynamic methods. Examples are Refs.~\cite{Mus92,
  Mus93}. These are approximate and purely numerical. (vi) In
addition, there are numerous papers that go beyond the model of
non-interacting Fermions and incorporate some aspects of the residual
interaction (see, for example, Refs.~\cite{Lau89, Alh93, Can94,
  Har98}).

Conceptually the present paper belongs to category (iii) although
technically it goes much beyond the earlier works. The tools
mentioned above make it possible to carry the approach analytically to
the very end, i.e., to the determination of the coefficients of the
orthogonal polynomials. All that is left to do is the numerical
evaluation of the formulas for a given set of parameters. In this way,
the approach provides analytical insight into the characteristic
dependence of the various level densities on energy and on the
parameters of the model. The approach shares the shortcomings of other
approaches in category (iii): it fails in the tails of the
spectrum. For the constant-spacing model this fact has been exhibited
in Ref.~\cite{Pal13}. We show here that similarly large discrepancies
arise for more realistic single-particle models if only low cumulants
are used to calculate the Fourier transform of $\rho_A(E)$. For the
construction of an approximation to $\rho_A(E)$ that is uniformly
valid throughout the spectrum, the use of cumulants of higher order
seems therefore indicated. We demonstrate why such a uniform
approximation, although theoretically desirable, is a practical
impossibility for the large values of single-particle states and
particle numbers of interest in this paper. We use the
constant-spacing model with a smooth single-particle level density
as an example and derive exact analytical expressions for the
expansion coefficients in terms of orthogonal polynomials. For $100$
particles in $200$ single-particle states, the density $\rho_A(E)$
takes values between unity and $\approx 10^{60}$, and the numerical
evaluation of these expressions would, therefore, require an accuracy
of one part in $10^{60}$ for a uniform approximation to $\rho_A(E)$.
This example shows why different approaches are needed in different
parts of the spectrum.

The paper is organized as follows. In Section~\ref{tot} we introduce
our method for the calculation of level densities and discuss its
limitations in the tails of the spectrum. This theoretical part is
followed in Section~\ref{examples} by a number of numerical results
for the three models of continuous single-particle level densities
mentioned above. In Section~\ref{asympt} we investigate the asymptotic
regime and justify the use of only the lowest moments and cumulants in
the Fourier transform. Particle-hole densities and the densities of
accessible states are derived in Sections~\ref{ph} and \ref{rhoacces},
respectively. The paper concludes with a summary and outlook.

\section{Total Level Density}
\label{tot}
\subsection{Introduction}

The total level density $\rho_A(E, J, \pi)$ is a function of energy
$E$, total spin $J$, and parity $\pi$. For $A$ non-interacting
Fermions in a spherical shell model, $\rho_A(E, J, \pi)$ has the
form~\cite{Blo68}
\be
\rho_A(E, J, \pi) = \frac{1}{2} \rho_A(E) \frac{2 J + 1}{2 \sqrt{2
\pi} \sigma^3} \exp \bigg\{ - \frac{(J + 1/2)^2}{2 \sigma^2} \bigg\}
\ .
\label{1}
\ee
Here $\sigma$ is the spin cutoff factor. We focus attention on
$\rho_A(E)$. By definition, this function (for brevity called ``the
level density'') is the density of levels versus energy $E$ obtained
by distributing $A$ Fermions over the states of the single-particle
model, each such state counted according to its multiplicity. We use a
single-particle model with a finite number $B$ of bound
single-particle states with energies $\ve_1 < \ve_2 < \ldots <
\ve_B$. For simplicity of notation we assume that these are not
degenerate. The fact that $B$ is finite strongly affects the energy
dependence of $\rho_A(E)$ at high excitation energies.

The eigenvalues $E_i$, $i = 1, 2, \ldots, N$ of the non-interacting
many-body system are obtained by distributing $A$ non-interacting
spinless Fermions over these states.  Here
\be
N = {B \choose A} \ .
\label{2}
\ee
We write the level density in the form
\be
\rho_A(E) = \sum_{i = 1}^N \delta(E - E_i) \ .
\label{5}
\ee
Eq.~(\ref{5}) shows that $\rho_A(E)$ is normalized to $N$. Explicitly,
$\rho_A(E)$ is given by (see Ref.~\cite{Boe70})
\be
\rho_A(E) = \sum_{\nu_1 = 0}^1 \sum_{\nu_2 = 0}^1 \ldots \sum_{\nu_B
= 0}^1 \delta_{\nu_1 + \nu_2 + \ldots + \nu_B, A} \ \delta(\nu_1 \ve_1
+ \nu_2 \ve_2 + \ldots + \nu_B \ve_B - E) \ .
\label{6}
\ee
For each single-particle state $j$ with $j = 1, \ldots, B$ the
Fermionic occupation number $\nu_j$ ranges from zero to one. The
Kronecker delta in Eq.~(\ref{6}) keeps the number of Fermions equal to
$A$. As in Eq.~(\ref{5}), the delta function is singular at every
eigenvalue $E_i$ of the non-interacting many-body system.

The energy $E_1$ of the non-degenerate ground state is
\be E_1 = \sum_{j = 1}^A \ve_j \ ,
\label{3}
\ee
and the energy of the highest state is
\be
E_N = \sum_{j = B + 1 - A}^B \ve_j \ .
\label{4}
\ee
We use the Fermi-gas model to calculate the mean energy $E_0$ of
$\rho_A(E)$. The occupation probability $n_{A, E}(\ve_j)$ of state
$j$ has the form
\be
n_{A, E}(\ve_j) = \frac{1}{1 + \exp \{ \beta \ve_j + \alpha \}} \ .
\label{7}
\ee
The parameters $\alpha$ and $\beta$ are obtained as solutions of the
equations
\be
A = \sum_j n_{A, E}(\ve_j) \ , \ E = \sum_j \ve_j n_{A, E}(\ve_j) \ ,
\label{8}
\ee
where $E$ is the total energy of the system. The value of $E_0$ is
obtained by setting $\beta = 0$ (infinite temperature),
\be
E_0 = \frac{A}{B} \sum_j \ve_j \ .
\label{9}
\ee
We consider a smoothed form $\overline{\rho}_A(E)$ of the level
density in Eqs.~(\ref{5}) and (\ref{6}). In units of the mean
single-particle level spacing, $\overline{\rho}_A(E)$ is of order
unity when $E$ is close to $E_1$ or $E_N$ while $\overline{\rho}_A(E)$
reaches values of the order $N$ in the center of the spectrum, i.e.,
for $E \approx E_0$. For medium-weight and heavy nuclei, $N$ as given
by Eq.~(\ref{2}) is a huge number easily attaining values like
$10^{30}$ or $10^{40}$. We aim at a reliable approximation to
$\overline{\rho}_A(E)$ that applies throughout most of the spectrum.

\subsection{Fourier Transform and Laplace Transform}

It is obviously difficult to deal with the level density in the form
of Eq.~(\ref{6}). An expression that is both manageable and amenable
to approximations is obtained by Fourier transformation with respect
to energy $E$ and by Laplace transformation with respect to particle
number $A$. To this end we write $\rho_A(E)$ in the form
\be
\rho_A(E) = {B \choose A} R_A(E) \ .
\label{10}
\ee
Then $R_A(E)$ is normalized to unity. To calculate the Fourier
transform ${\cal F}_A(\tau)$ of $R_A(E)$ we define
\be
\tilde{\ve}_j = \ve_j - \Delta\ ,
\label{11}
\ee
where
\be
\Delta = \frac{1}{B} \sum_{j = 1}^B \ve_j = \frac{E_0}{A}
\label{12}
\ee
with $E_0$ defined in Eq.~(\ref{9}). Then
\be
R_A(E) = \frac{1}{N} \sum_{\nu_1, \nu_2, \ldots, \nu_B = 0}^1
\delta_{\nu_1 + \nu_2 + \ldots + \nu_B, A} \delta(\nu_1 \tilde{\ve}_1 + \nu_2
\tilde{\ve}_2 + \ldots + \nu_B \tilde{\ve}_B - (E - E_0) ) \ .
\label{13}
\ee
The Fourier transform ${\cal F}_A(\tau)$ of $R_A(E)$ is given by
\ba
{\cal F}_A(\tau) &=& \int_{- \infty}^{+ \infty} {\rm d} E \ \exp \{ i
(E - E_0) \tau \} R_A(E) \nonumber \\
&=& \frac{1}{N}  \sum_{\nu_1, \nu_2, \ldots, \nu_B = 0}^1 \delta_{\nu_1
+ \nu_2 + \ldots + \nu_B, A} \exp \bigg\{ i \tau \sum_{j = 1}^B \nu_j
\tilde{\ve}_j \bigg\} \ .
\label{14}
\ea
The Laplace transform of ${\cal F}_A(\tau)$ with respect to $A$ is
\ba
L(\alpha, \tau) &=& \sum_A {\cal F}_A(\tau) \exp \{ \alpha A \}
\nonumber \\
&=& \frac{1}{N} \sum_{\nu_1 = 0}^1 \sum_{\nu_2 = 0}^1 \ldots
\sum_{\nu_B = 0}^1 \exp \bigg\{ \sum_{j = 1}^B \nu_j ( \alpha + i \tau
\tilde{\ve}_j ) \bigg\} \nonumber \\
&=& \frac{1}{N}  \prod_{j = 1}^B (1 + \exp \{ \alpha + i \tau
\tilde{\ve}_j \} ) \ .  
\label{15}
\ea
We note that ${\cal F}_A(\tau)$ is the coefficient multiplying $\exp
\{ \alpha A \}$ in a Taylor series expansion of $L(\alpha, \tau)$ in
powers of $\exp \{ \alpha \}$.

\subsection{Moments and Cumulants}

The closed-form expression~(\ref{15}) allows us to calculate explicit
expressions for the low moments and low cumulants of
$\rho_A(E)$. These are then used to construct approximations to
$\rho_A(E)$. For $k = 0, 1, \ldots$ we define the $k^{\rm th}$ moments
$m_A(k)$ and the normalized moments $M_A(k)$ of $\rho_A(E)$ by
\ba
m_A(k) &=& \int {\rm d} E \ (E - E_0)^k \rho_A(E) \ , \nonumber \\
M_A(k) &=& \int {\rm d} E \ (E - E_0)^k R_A(E) \nonumber \\ 
&=& \frac{1}{i^k} \frac{\partial^k}{\partial \tau^k} {\cal F}_A(\tau) 
\bigg|_{\tau = 0} \ . 
\label{16}
\ea
For the lowest moments we proceed as in Ref.~\cite{Pal13}. We write $N
L(\alpha, \tau) = \exp \{ H(\alpha, \tau) \}$ and expand $H(\alpha,
\tau)$ in a Taylor series around $\tau = 0$. We define
\be
f(\alpha) = \frac{\exp \{ \alpha \}}{1 + \exp \{ \alpha \}}
\label{17}
\ee
and denote by $f^{(n)}$ the $n^{\rm th}$ derivative of $f$. Since $f'
= f - f^2$, the $n^{th}$ derivative $f^{(n)}$ has the form
\be
f^{(n)} = \sum_{l = 1}^{n + 1} c^{(n)}_l f^l
\label{19}
\ee
with integer coefficients $c^{(n)}_l$. We list the first few such
forms,
\ba
f' &=& f (1 - f) \ , \nonumber \\
f'' &=& f (1 - f) (1 - 2 f) \ , \nonumber \\
f''' &=& f (1  - f) (1 - 6 f + 6 f^2) \ , \nonumber \\
f^{(4)} &=& f (1 - f) (1 - 14 f + 36 f^2 - 24 f^3) \ , \nonumber \\ 
f^{(5)} &=& f (1 - f) (1 - 30 f + 150 f^2 - 240 f^3 + 120 f^4) \ .
\label{20}
\ea
It is straightforward to generate the expressions for the
higher-order derivatives. With these definitions we have for $k = 1,
2, \ldots$
\be
\frac{\partial^k}{\partial \tau^k} H(\alpha, \tau) \bigg|_{\tau = 0} =
f^{(k - 1)} \ i^k \sum_{j = 1}^B (\tilde{\ve}_j)^k \ .
\label{18}
\ee
The definition~(\ref{11}) implies $\sum_j \tilde{\ve}_j = 0$. Using
the Taylor expansion for $H$ we obtain
\be
L(\alpha, \tau) = \frac{1}{N} (1 + \exp \{ \alpha \})^B \exp \bigg\{
\sum_{k = 2}^\infty \frac{i^k \tau^k}{k!} f^{(k - 1)} \sum_{j = 1}^B
(\tilde{\ve}_j)^k \bigg\} \ .
\label{21}
\ee
According to Eqs.~(\ref{14}) and (\ref{15}), the moments $m_A(k)$ of
$\rho_A(E)$ are given by the coefficients multiplying $\exp \{ \alpha
A \}$ in a Taylor series in powers of $\exp \{ \alpha \}$ of the
$k^{th}$ derivative of $N L(\alpha, \tau)$ taken with respect to $i
\tau$ at $\tau = 0$. For $k = 0, 1, \ldots, 6$ and $k = 8$ these are
listed in \ref{momentsApp}. We note that the moments $m_A(k)$ depend only on
the moments $\sum_{j = 1}^B (\tilde{\ve}_j)^l$ with $l \leq k$ of the
single-particle level density. Moreover, the even (odd) moments are
even (odd, respectively) with respect to the interchange $A
\leftrightarrow (B - A)$, as required by particle-hole symmetry.

Using $M_A(k) = m_A(k) / N$ and the last of Eqs.~(\ref{16}) we have
\be
{\cal F}_A(\tau) = \sum_{k = 0}^\infty \frac{i^k \tau^k}{k!} M_A(k) \ .
\label{23}
\ee
With $M_A(0) = 1$ and $M_A(1) = 0$ this can be written as
\ba
{\cal F}_A(\tau) &=& \exp \bigg\{ \ \ln \bigg( 1 + \sum_{k = 2}^\infty
\frac{i^k \tau^k}{k!} M_A(k) \bigg) \bigg\} \nonumber \\
&=& \exp \bigg\{ \sum_{k = 2}^\infty \frac{i^k \tau^k}{k!} \kappa_A(k)
\bigg\} \ .
\label{24}
\ea
The cumulants $\kappa_A(k)$ are polynomial expressions in the moments
$M_A(k')$ with $k' \leq k$. The lowest cumulants are given by
\ba
\kappa_A(2) &=& M_A(2) \ , \nonumber \\
\kappa_A(3) &=& M_A(3) \ , \nonumber \\
\kappa_A(4) &=& M_A(4) - 3 M^2_A(2) \ , \nonumber \\
\kappa_A(5) &=& M_A(5) - 10 M_A(2) M_A(3) \ , \nonumber \\
\kappa_A(6) &=& M_A(6) - 15 M_A(2) M_A(4) - 10 M^2_A(3) + 30 M_A^3(2)
\ .
\label{25}
\ea
For the constant-spacing model we have $\kappa_A(7) = 0$ and
\ba
\kappa_A(8) &=& M_A(8)-28 M_A(2) M_A(6)-35 M^2_A(4)+420 M^2_A(2) M_A(4)
\nonumber \\
&& - 630 M^4_A(2)\ .
\label{25a}
\ea

\subsection{Orthogonal Polynomials}

An approximate expression for the Fourier transform ${\cal F}_A(\tau)$
is obtained by inserting a number of the lowest cumulants
$\kappa_A(k)$ into Eq.~(\ref{24}). Calculating from here the function
$R_A(E)$ by inverting the Fourier transformation~(\ref{14}) is
numerically cumbersome. The integration over $\tau$ can be avoided by
using orthogonal polynomials.

We recall that the function $R_A(E)$ defined in Eq.~(\ref{10}) is
normalized to unity. Moreover, $R_A(E)$ differs from zero only in the
interval ${\cal I} = \{ E_1, E_N \}$. At the end points of that
interval, $R_A(E)$ has the value $1 / N \approx 0$. It is, therefore,
meaningful to expand $R_A(E)$ in the interval ${\cal I}$ in terms of
orthonormal polynomials that vanish at the end points. For a variable
$x$ defined in the interval $- 1 \leq x \leq + 1$, such polynomials
are
\ba
T_n &=& \sin \{ (\pi/2) n x \} \ {\rm for} \ n \ {\rm positive \ and \
even} \ , \nonumber \\
T_n &=& \cos \{ (\pi/2) n x \} \ {\rm for} \ n \ {\rm positive \ and \
odd} \ .
\label{26}
\ea
We define
\be
L = E_N - E_1 \ , \ E_c = \frac{1}{2}(E_1 + E_N)\ ,
\label{27}
\ee
where the index $c$ stands for center. Writing $E = E_c + L x / 2$ we
map the interval $- 1 \leq x \leq 1$ onto the interval ${\cal I}$.
For the polynomials $T_n$ that map yields
\ba
T_n(E) &=& \sqrt{\frac{2}{L}} \sin \{ \frac{1}{L} \pi n (E - E_c) \} \
{\rm for} \ n \ {\rm positive \ and \ even} \ , \nonumber \\
T_n(E) &=& \sqrt{\frac{2}{L}} \cos \{ \frac{1}{L} \pi n (E - E_c) \} \
{\rm for} \ n \ {\rm positive \ and \ odd} \ .
\label{28}
\ea
We expand
\be
R_A(E) = \sum_{n = 1}^\infty r_A(n) T_n(E) \ .
\label{29}
\ee
Then
\be
r_A(n) = \int_{E_1}^{E_N} {\rm d} E \ R_A(E) T_n(E) \ .
\label{30}
\ee
Writing the functions $T_n$ in Eqs.~(\ref{28}) as superpositions of
$\exp \{ \pm i \pi n (E - E_c) / L \}$ and using the first of
Eqs.~(\ref{14}), we find that the coefficients $r_A(n)$ are given by
\ba
r_A(n) &=& \frac{1}{i} \sqrt{\frac{1}{2 L}} \exp \{ i \pi n (E_0 -
E_c) / L \} {\cal F}_A(\pi n / L) + c.c. \nonumber \\
&& \qquad \qquad \ {\rm for} \ n \ {\rm positive \ and \ even} ,
\nonumber \\
r_A(n) &=& \sqrt{\frac{1}{2 L}} \exp \{ i \pi n (E_0 - E_c) / L \}
{\cal F}_A(\pi n / L) + c.c. \nonumber \\
&& \qquad \qquad \ {\rm for} \ n \ {\rm positive \ and \ odd}  .
\label{31}
\ea
%

\subsection{Smooth Single-Particle Level Density}
\label{sing}

Eq.~(\ref{21}) shows that $\rho_A(E)$ is completely determined by the
moments of the single-particle energies $\ve_1 < \ve_2 < \ldots <
\ve_B$. The moments may be worked out for any single-particle level
density given in the form
\be
\rho_1(\ve) = \sum_{j = 1}^B \delta(\ve - \ve_j) \ .
\label{32}
\ee
For what follows it is convenient to consider a smooth
single-particle level density $\overline{\rho}_1(\ve)$ rather than a
sum of delta functions as in Eq.~(\ref{32}). The function
$\overline{\rho}_1(\ve)$ is defined for $\ve$ in the interval $0 \leq
\ve \leq V$. The letter $V$ is chosen as a reminder of the depth of
the single-particle potential. In terms of $\overline{\rho}_1(\ve)$,
the single-particle energies $\ve_j$ are obtained as solutions of the
equations
\be
\int_0^{\ve_j} {\rm d} \ve \ \overline{\rho}_1(\ve) = j \ , \ j = 1,
\ldots, B \ .
\label{33}
\ee
Here $B$ is the maximum integer for which the solution $\ve_B$ of
Eq.~(\ref{33}) obeys $\ve_B < V$. With $A$ non-interacting spinless
Fermions distributed over $B$ single-particle states, the total
number $N$ of many-body states is given by Eq.~(\ref{2}). The Fermi
energy $F$ for $A$ Fermions is defined by
\be
\int_0^{F} {\rm d} \ve \ \overline{\rho}_1(\ve) = A \ .
\label{34}
\ee
The spectrum of $\rho_A(E)$ ranges from $E_1$ to $E_N$ where
\be
E_1 = \int_0^F {\rm d} \ve \ \ve \ \overline{\rho}_1(\ve) \ , \
E_N = \int_{E_{\rm sup}}^V {\rm d} \ve \ \ve \ \overline{\rho}_1(\ve) \ ,
\label{35}
\ee
with $E_{\rm sup}$ defined by
\be
A = \int_{E_{\rm sup}}^{V} {\rm d} \ve \ \overline{\rho}_1(\ve) \ .
\label{36}
\ee
Eqs.~(\ref{7}) to (\ref{9}) take the form
\be
n_{A, E}(\ve) = \frac{1}{1 + \exp \{ \beta \ve + \alpha \}}
\label{37}
\ee
where the parameters $\alpha$ and $\beta$ are solutions of the
equations
\be
A = \int_0^V {\rm d} \ve \ n_{A, E}(\ve) \overline{\rho}_1(\ve) \ ,
\ E = \int_0^V {\rm d} \ve \  \ve n_{A, E}(\ve)
\overline{\rho}_1(\ve) \ .
\label{38}
\ee
The value of $E_0$ is obtained by setting $\beta = 0$,
\be
E_0 = \frac{A}{B} \int_0^V {\rm d} \ve \  \ve \overline{\rho}_1(\ve)
\ .
\label{39}
\ee
The moments of the single-particle energies are simply given by
\be
\sum_j \ve^k_j = \int_0^V {\rm d} \ve \ \ve^k \rho_1(\ve) \ .  
\label{39a}
\ee
Thus, all ingredients for calculating the  moments and
cumulants for the level densities are available. 

Three numerical examples for a smooth single-particle level density 
are presented in Section~\ref{examples} below. The results show that our
approximation scheme for the many-body level density $\rho_A(E)$
works well within an energy interval centered in the middle of the
spectrum and covering about half the total range. The scheme fails in
the tails of $\rho_A(E)$. This may appear as an unsatisfactory aspect
of our work. Indeed, it would be highly desirable to develop an
approximation to $\rho_A(E)$ that is uniformly reliable throughout the
entire spectrum. In \ref{searchApp} we show why for $B \gg 1$ and $A \gg 1$
this aim is beyond reach.

An alternative to a uniform approximation for the level density
consists in using different approaches to different parts of the
spectrum of $\rho_A(\varepsilon)$. It is shown below that the approach
developed in this paper is accurate for a range of energies where
$\rho_A(\varepsilon) \gtrsim 10^{- 20} {B \choose A}$. For values of
$\rho_A(E) \leq 10^{10}$ or $10^{15}$ most of the approaches mentioned
in the Introduction can be used. In the intermittent energy domain one
must probably resort to some sort of interpolation.

\section{Examples}
\label{examples}

In Ref.~\cite{Pal13} the level density $\rho_A(E)$ was calculated
approximately for a single-particle model with constant level spacing
$d$, $\rho_1(E) = \sum_j \delta(E - d j)$. For medium-weight and
heavy nuclei, that model is unrealistic. The approach presented in
this paper allows for a single-particle spectrum with an arbitrary
sequence of single-particle energies. We demonstrate some results of
this generalization. We use a continuous single-particle level
density as in Section~\ref{sing} and consider three choices of
$\overline{\rho}_1(\ve)$,
\ba
\overline{\rho}^{(0)}_1(\ve) &=& \frac{A}{F} \ , \nonumber \\
\overline{\rho}^{(1)}_1(\ve) &=& \frac{2 A}{F^2} \ve \ , \nonumber \\
\overline{\rho}^{(2)}_1(\ve) &=& \frac{3 A}{F^3} \ve^2 \ . 
\label{40}
\ea
The normalization constants are determined by Eq.~(\ref{34}). The
constant-spacing model ($\overline{\rho}^{(0)}_1(\ve)$) is considered
for the sake of comparison with Ref.~\cite{Pal13}. A linear
(quadratic) dependence of $\overline{\rho}_1(\ve)$ on energy
approximates the single-particle spectrum in medium-weight (in
heavy) nuclei, respectively. However, the three cases considered in
Eqs.~(\ref{40}) serve as examples only. More realistic forms of the
smooth single-particle level density $\overline{\rho}_1(\ve)$ have
been constructed, see, for instance, Ref.~\cite{Shl92}. These show a
strong rise of $\overline{\rho}_1(\ve)$ versus $\ve$ up to $\ve = 0$
(zero binding energy), followed by a sharp drop. Such models can
easily be used in our context, the moments of the single-particle
energies being given by Eq.~(\ref{39a}).

The input parameters of the model~(\ref{40}) are the range $V$ of
the single-particle spectrum, the Fermi energy $F$, the number $A$ of
Fermions, and the power of the energy dependence of the
single-particle level density $\overline{\rho}_1(E)$. The total
numbers $B^{(0)}$, $B^{(1)}$ and $B^{(2)}$ of single-particle states
are given by
\be
B^{(0)} = A \frac{V}{F} \ , ß B^{(1)} = A \frac{V^2}{F^2} \ , \ B^{(2)} =
A \frac{V^3}{F^3} \ .
\label{41}
\ee
Eqs.~(\ref{34}), (\ref{35}), and (\ref{36}) yield
\ba
\begin{matrix} E^{(0)}_1 = \frac{1}{2} A F \ , & E^{(1)}_1 = \frac{2}{3}
A F \ , & E^{(2)}_1 = \frac{3}{4} A F \ , \cr
E^{(0)}_0 = \frac{1}{2} A V \ , & E^{(1)}_0 = \frac{2}{3} A V \ , &
E^{(2)}_0 = \frac{3}{4} A V \ , \cr
E^{(0)}_{\rm sup} = V - F \ , & (E^{(1)}_{\rm sup})^2 = V^2 - F^2 \ ,
& (E^{(2)}_{\rm sup})^3 = V^3 - F^3 \ , \cr
E^{(0)}_N = \frac{1}{2} A \frac{V^2 - (E^{(0)}_{\rm sup})^2}{F} \ , 
& E^{(1)}_N = \frac{2}{3} A \frac{V^3 - (E^{(1)}_{\rm sup})^3}{F^2} \ ,
& E^{(2)}_N = \frac{3}{4} A \frac{V^4 - (E^{(2)}_{\rm sup})^4}{F^3} \ .
\cr
\end{matrix}
\label{42}
\ea
We note that for fixed $A$ and $V$ and with increasing power of $\ve$
governing $\overline{\rho}_1(\ve)$, the spectrum shifts towards higher
energies.

For the three cases defined in Eq.~(\ref{40}), the moments of the
single-particle energies required for the calculation of moments and
cumulants of $\rho_A(E)$ can easily be worked out and are given in
\ref{singpartApp}. We note that the odd moments vanish for the case of
constant single-particle level spacing and are negative in the other
two cases. This is expected since in this case the distance between
neighboring single-particle states decreases with increasing $j$, see
Eqs.~(\ref{40}).

\subsection{Numerical Results}

Inserting Eqs.~(\ref{47}) into Eqs.~(\ref{22}) we obtain explicit
expressions for the moments, from here for the cumulants in
Eqs.~(\ref{25}), for the Fourier transform in Eq.~(\ref{24}), and for
the coefficients $r_A(n)$ in Eqs.~(\ref{31}). These are used to
generate the numerical results of the present Section.

\begin{figure}[ht]
\vspace{5 mm}
\includegraphics[width=0.8\linewidth]{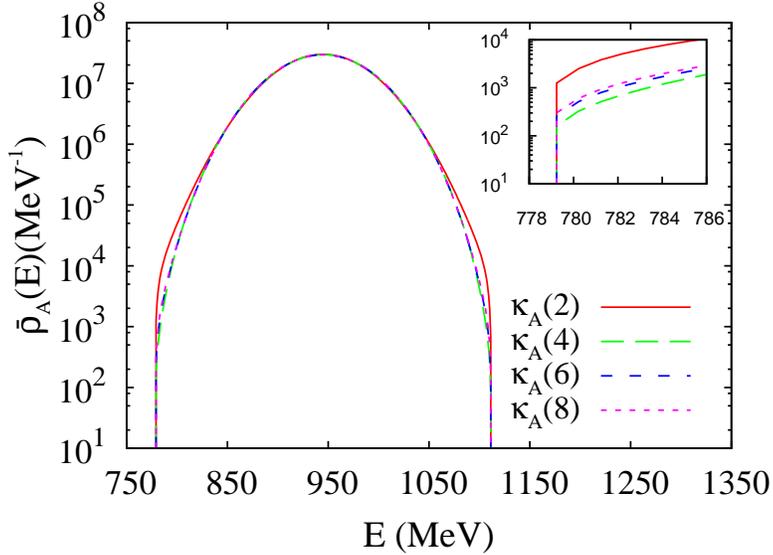}
\vspace{3 mm}
\caption{Level density $\overline{\rho}_A(E)$ as a function of energy
  $E$ for the case of $\overline{\rho}^{(0)}_1$ and for parameter
  values given in the text. The cumulant sum of the Fourier
  transform~(\ref{24}) includes the terms up to the listed cumulant
  $\kappa_A$. The inset shows $\overline{\rho}_A(E)$ in the tails.}
\label{fig1}
\end{figure}

We first consider the case of constant single-particle level spacing
that has also been investigated in Ref.~\cite{Pal13}. For sufficiently
small values of the particle number $A$ and level number $B$, exact
values for the total level density are available for comparison. For
constant spacing, all odd moments and cumulants vanish and the
distribution is symmetric about the center of the spectrum.
Fig.~\ref{fig1} presents the level density $\overline{\rho}_A(E)$ for
$A = 42$ and $B^{(0)} = 51$ for different cutoffs of the cumulant sum
in the Fourier transform~(\ref{24}). Further parameters are the Fermi
energy $F = 37$~MeV and the range of the single-particle spectrum
$V = 45$~MeV. The values of $A$, $B^{(0)}$, $F$ and $V$ are consistent
with Eq.~(\ref{41}) for constant spacing. The many-body spectrum
extends from $E_1 = (1/2) AF = 778$ MeV to $E_N = (A/2) (V^2 - (V -
F)^2)/F = 1112$ MeV.  We find that for these parameters the use of $n
= 30$ orthogonal polynomials is sufficient; no significant changes occur as $n$ is
increased further.  Increasing the number of cumulants in the Fourier
transform sum (\ref{24}) produces changes only in the tails of the
level density, as can be seen in the inset of Fig.~\ref{fig1}.

\begin{figure}[ht]
\vspace{5 mm}
\includegraphics[width=0.8\linewidth]{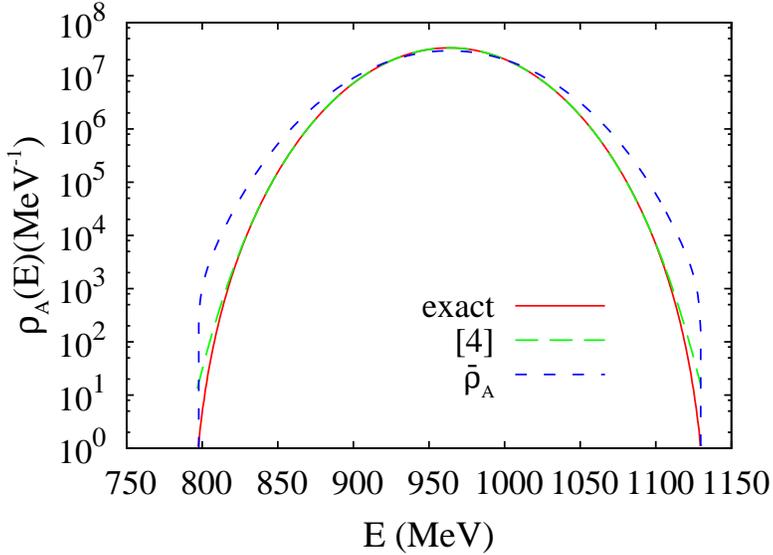}
\vspace{3 mm}
\caption{Comparison of our present results for the continuous
  single-particle level density $\overline{\rho}_1^{(0)}$ using
  cumulants up to $\kappa_A(8)$ (short dashed blue line) with the
  exact level density $\rho_A(E)$ (red solid line) and results of the
  fitting procedure described in Ref.~\cite{Pal13} (long dashed green
  line) for $A = 42$ and $B^{(0)} = 51$. For a better visualization
  the level density $\overline{\rho}_A$ is shifted by $(d/2) A = 18$
  MeV towards higher energies.}
\label{fig2}
\end{figure}

We compare results for the continuous single-particle level density
given by the first of Eqs.~(\ref{40}) with those of Ref.~\cite{Pal13}
where a sum of delta functions was used. In Ref.~\cite{Pal13}, the
level density $\overline{\rho}_A(E)$ is calculated exactly for
sufficiently small values for $B$ and $A$ and, in addition, is
approximated by fitting the expression $\overline{\rho}_A(E) \propto
\exp \{ - \gamma_2 (E)^2 - \gamma_4 (E)^4 - \gamma_6 (E)^6 \}$ to the
first three moments $M_A(2)$, $M_A(4)$, $M_A(6)$. For the comparison
we consider for both exact and fitted results \cite{Pal13} the
constant spacing $d=V/B^{(0)}=0.88$ MeV. We use the first four
non-vanishing cumulants $\kappa_A(2)$, $\kappa_A(4)$, $\kappa_A(6)$
and $\kappa_A(8)$ for the Fourier transform in Eq.~(\ref{24}) and $n =
30$ orthogonal polynomials. In the present case, the spectrum extends
from $(d/2) A^2$ to $d [B^{(0)} A - (1/2) A^2]$ while in the case of
Ref.~\cite{Pal13} it extends from $(d/2) A (A + 1)$ to $d [B^{(0)} A -
  (1/2) A (A - 1)]$. We have taken account of this difference by
shifting the spectrum calculated in the present framework by $+(d/2)
A$. For the same set of parameters as in Fig.~\ref{fig1},
Fig.~\ref{fig2} shows the exact values for the level density
$\overline{\rho}_A(E)$ (solid red line), the approximation of
Ref.~\cite{Pal13} based upon a fit using up to the sixth moment (long dashed
green line), and the approximation using up to the eighth
cumulant and the expansion in terms of orthogonal polynomials (short
dashed blue line, shifted). We see that both approximate methods,
while very good near the center of the spectrum, fail near the
boundaries where $\overline{\rho}_A(E) = 1$. By construction, the
method of orthogonal polynomials yields $\overline{\rho}_A(E) = 0$ at
the minimum and maximum energies $E_1$ and $E_N$ of the
spectrum. Close to the spectrum tails, however, the spectrum is
broader than the exact results and overestimates the exact values
somewhat more than the fitting procedure of Ref.~\cite{Pal13}.

\begin{figure}[ht]
\vspace{5 mm}
\includegraphics[width=0.8\linewidth]{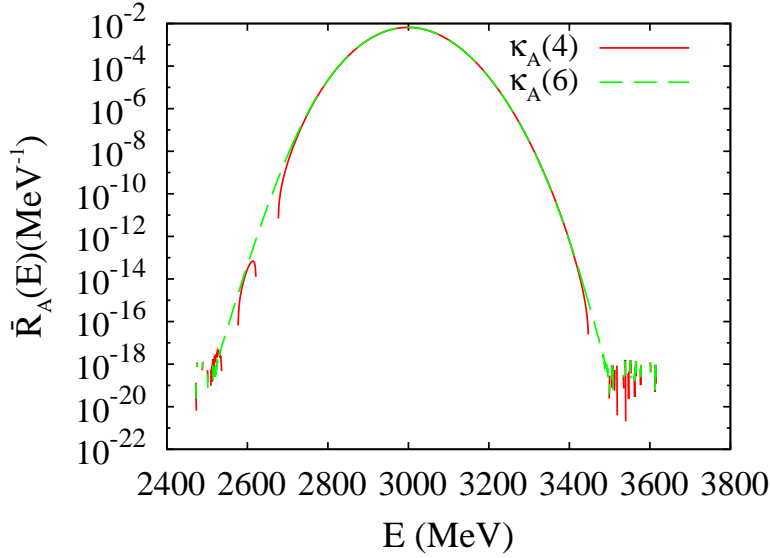}
\vspace{3 mm}
\caption{Normalized level density $\overline{R}_A(E)$ as a function of
  energy $E$ for $\overline{\rho}^{(1)}_1(\ve)$ and $A = 100$ and
  $B^{(1)} = 148$. The Fourier transform is calculated according to
  Eq.~(\ref{24}) using the cumulants of Eqs.~(\ref{25}) up to
  $\kappa_A(4)$ (red solid line) and $\kappa_A(6)$ (green dashed line).}
\label{fig3}
\end{figure}

Compared to Ref.~\cite{Pal13}, the strength of our method is that
we can go beyond the constant-spacing model. For medium-weight (heavy)
nuclei, we use the linear (quadratic) single-particle level density
$\overline{\rho}^{(1)}_1(\ve)$ ($\overline{\rho}^{(2)}_1(\ve)$) of
Eq.~(\ref{40}), respectively. Here the odd moments and cumulants of
the distribution also come into play, and the many-body density is no
longer symmetric about its center. For $\overline{\rho}^{(1)}_1(\ve)$,
Fig.~\ref{fig3} shows the normalized level density $\overline{R}_A(E)$
as a function of energy $E$ for $A = 100$ and $B^{(1)} = 148$ and two
different cutoffs in the Fourier transform sum~(\ref{24}). The Fermi
energy and the range of the single-particle spectrum are $F = 37$~MeV
and $V = 45$~MeV, respectively. The many-body spectrum extends from
$E_1 = $2466 MeV to $E_N = $3620 MeV. We find that for these
parameters the use of $n = 50$ orthogonal polynomials is sufficient;
no noticeable changes occur as $n$ is increased further. For the level density
calculated using cumulants up to $\kappa_A(4)$ in Eq.~(\ref{24}) there
exist intervals in the tails of the spectrum where $\overline{R}_A(E)$
becomes negative (and is therefore not displayed on the logarithmic
scale), while using cumulants up to $\kappa_A(6)$ in Eq.~(\ref{24})
yields strictly positive values.

\begin{figure}[ht]
\vspace{5 mm}
\includegraphics[width=0.8\linewidth]{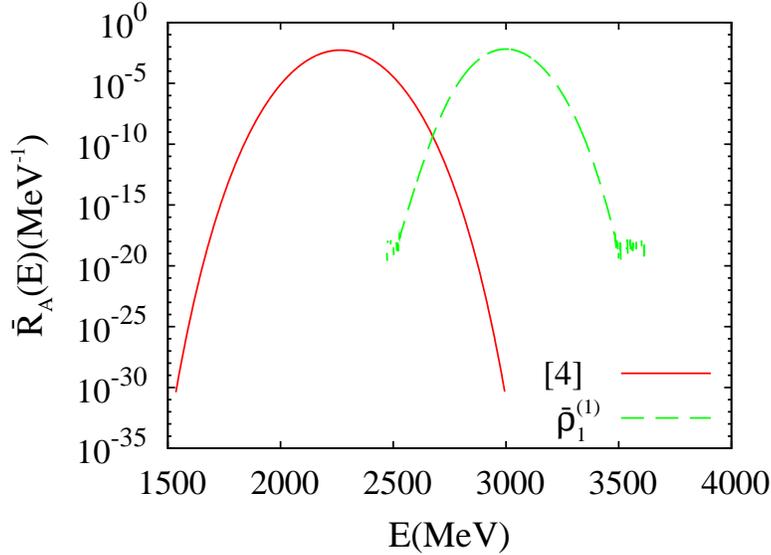}
\vspace{3 mm}
\caption{The normalized level density $\overline{R}_A(E)$ for
  $\overline{\rho}^{(1)}_1$ and $A = 100$, $B^{(1)} = 148$ considering
  cumulants up to $\kappa_A(6)$ (green dashed line) compared to the
  result of the constant-spacing model with single-particle level
  spacing $d = V / B^{(1)}$ using the fitting procedure described in
  Ref.~\cite{Pal13} (full red line).}
\label{fig4}
\end{figure}

In Fig.~\ref{fig4} we compare our result using
$\overline{\rho}^{(1)}_1$ and cumulants up to $\kappa_A(6)$ with the
normalized level density for the constant-spacing model calculated as
described in Ref.~\cite{Pal13}, with single-particle level spacing $d
= V / B^{(1)}$ and $V = 45$ MeV. The latter method yields values for
the level density $\overline{R}_A(E)$ throughout the spectrum but
overestimates $\overline{R}_A(E)$ in the tails while our present
result does not cover values of $\overline{R}_A(E)$ that are smaller
than $10^{-21}$ times the maximum. We note three significant
differences between the results for the constant-spacing model and for
$\overline{\rho}^{(1)}_1$. (i) The spectrum is shifted by 700 MeV
towards higher energies. However, part of this shift is due to an
increase of the ground-state energy. (ii) The width of the many-body
level density is smaller for $\overline{\rho}^{(1)}_1$ than for the
constant-spacing model. (iii) For $\overline{\rho}^{(1)}_1$ (and also
for $\overline{\rho}^{(2)}_1$), the odd cumulants $\kappa_A(3) < 0$
and $\kappa_A(5) < 0$ cause an asymmetry in the many-body level
density. Therefore, the maximum of $\overline{R}_A(E)$ is below the
center of the spectrum in both cases. For $\overline{\rho}^{(1)}_1$ it
occurs at $3000$~MeV while the center is at $E_c = 3043$~MeV. The
temperature of the system, defined in terms of the inverse of the
derivative of $\overline{\rho}_A(E)$, becomes infinite at the maximum.
This is consistent with the Fermi-gas model of Eqs.~(\ref{5}) to
(\ref{7}) where infinite temperature (defined by $\beta = 0$) is
attained at the mean energy $E_0 = 3000$ MeV of the system.

\begin{figure}[ht]
\vspace{5 mm}
\includegraphics[width=0.8\linewidth]{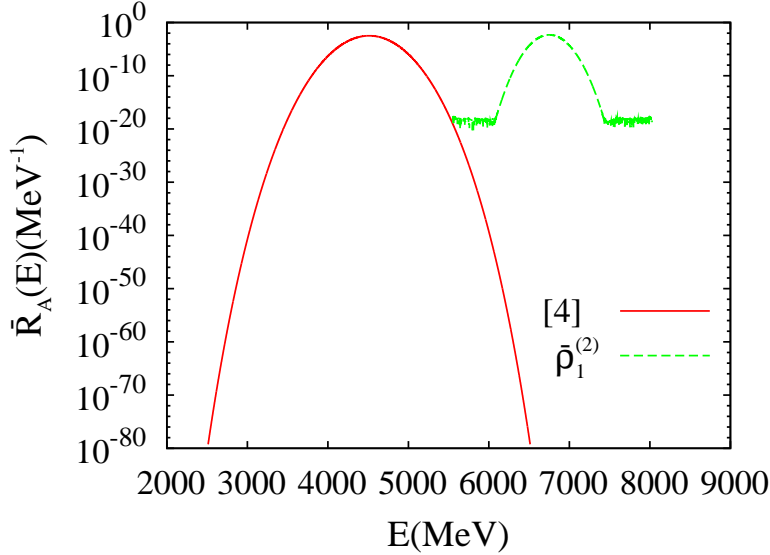}
\vspace{3 mm}
\caption{The normalized level density $\overline{R}_A(E)$ for
  $\overline{\rho}^{(2)}_1$ and $A = 200$, $B^{(2)} = 360$ considering
  cumulants up to $\kappa_A(6)$ (dashed green line) compared to
  results of the constant-spacing model with single-particle level
  spacing $d = V / B^{(2)}$ using the method described in
  Ref.~\cite{Pal13} (solid red line).}
\label{fig5}
\end{figure}

For the case of $\overline{\rho}^{(2)}_1$ we consider $A = 200$,
$B^{(2)} = 360$, $V = 45$ MeV and $F = 37$ MeV and calculate the
normalized level density $\overline{R}_A(E)$ taking into account the
cumulants in Eqs.~(\ref{25}) up to $\kappa_A(6)$. The number of
orthogonal polynomials used for the calculation is $n = 100$. The ratio $\overline{\rho}_A(E) /
\overline{R}_A(E) \approx 10^{106}$ is huge. A comparison with the
density for constant single-particle level spacing $d = V / B^{(2)}$
calculated as in Ref.~\cite{Pal13} is displayed in Fig.~\ref{fig5} and
shows the same features as seen for $\overline{\rho}^{(1)}_1$: the
distribution is shifted towards higher energies, is more narrow, and
is asymmetric. As for $\overline{\rho}^{(1)}_1$, our approximation
covers roughly the leading $20$ orders of magnitude of the
distribution.

%

\section{Asymptotic Expansion \label{asympt}}

The results in Section~\ref{examples} are obtained by using only the
lowest cumulants in the Fourier transform of Eq.~(\ref{24}). The
omission of higher-order cumulants can be justified asymptotically,
i.e., for $B \gg 1$ and $A \gg 1$. (Here and in what follows $B, A \gg
1$ means $A \approx B/2 \gg 1$. Because of particle-hole symmetry, the
asymptotic approximation fails not only as $A$ approaches $1$ from
above but also as $A$ approaches $B$ from below). In Ref.~\cite{Pal13}
we have shown for the constant-spacing model that in the asymptotic
regime the contributions to the sum over $k$ in Eq.~(\ref{21})
decrease very rapidly with increasing $k$. In \ref{asympApp} we extent
this proof to the two other cases in Eqs.~(\ref{40}). In both these
cases, too, the rescaled cumulants decrease very rapidly with
increasing $k$. As a by-product we find that the resulting asymptotic
expression for the Fourier transform in Eq.~(\ref{24}) becomes much
simpler than the full expression.

We emphasize that the asymptotic behavior of the rescaled cumulants
obtained in Eqs.~(\ref{55}) is generic. Indeed, this behavior is due
to the fact that the unscaled cumulants are all proportional to
$A$. This fact, in turn, is due to the normalization
condition~(\ref{34}). For an arbitrary single-particle level density
$\omega_1(\ve)$ that same condition yields for the normalized form
$\overline{\rho}_1(E) = A \omega_1(\ve) / \int_0^F {\rm d} \ve'
\omega(\ve')$. The moments of single-particle energies in
Eqs.~(\ref{47}) can generally be written as $\sum_j (\ve_j)^k =
\int_0^V {\rm d} \ve \ \overline{\rho}_1(\ve) \ve^k$. Thus, all these
moments and the corresponding cumulants are proportional to $A$. This
implies the asymptotic form~(\ref{55}). The constants multiplying
$A^{-k/2}$ depend on the form of $\omega(\ve)$.  We expect that these
constants are of order unity for any function $\omega(\ve)$ that
distributes the single-particle energies $\ve_j$ more or less
uniformly over the interval $\{0, V\}$.

\subsection{Examples}

\begin{figure}[ht]
\vspace{5 mm}
\includegraphics[width=0.8\linewidth]{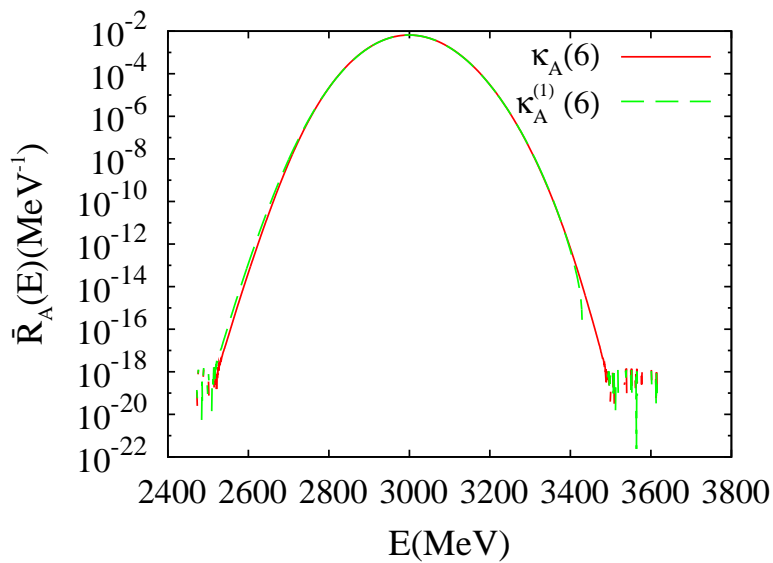}
\vspace{3 mm}
\caption{The normalized level density $\overline{R}_A(E)$ for
  $\overline{\rho}_1^{(1)}(E)$ and $A = 100$, $B^{(1)} = 148$
  using the first six exact (red solid line) and asymptotic
  (green dashed line) cumulants. The number of orthogonal polynomials used for the calculation
   is $n=50$.}
\label{fig6}
\end{figure}

\begin{figure}[ht]
\vspace{5 mm}
\includegraphics[width=0.8\linewidth]{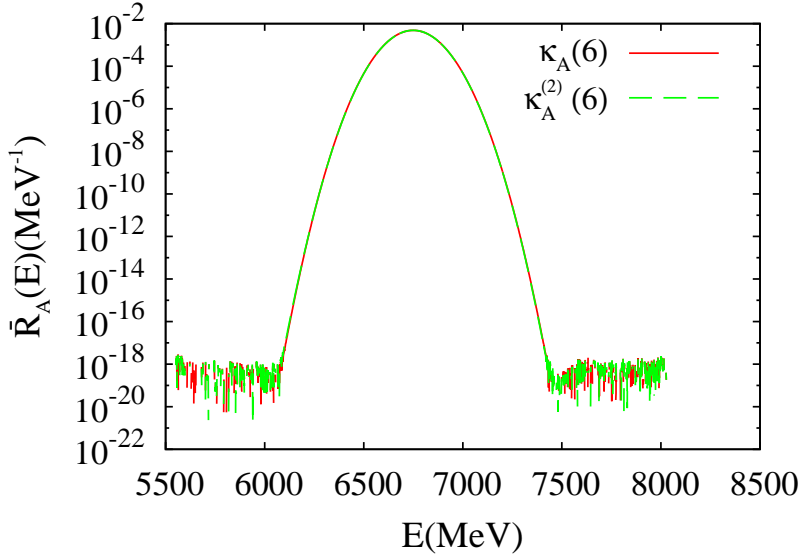}
\vspace{3 mm}
\caption{The normalized level density $\overline{R}_A(E)$ for
  $\overline{\rho}_1^{(2)}(E)$ and $A = 200$, $B^{(2)} = 360$ using
  the first six exact (red solid line) and asymptotic (green dashed
  line) cumulants. The number of orthogonal polynomials used for the calculation is
  $n=100$.}
\label{fig7}
\end{figure}

For a comparison of the asymptotic with the full results we need the
asymptotic values of the cumulants. Eq.~(\ref{41}) gives $A / B = F^2
/ V^2$ ($A / B = F^3 / V^3$) for $\rho^{(1)}_1$ ($\rho^{(2)}_1$,
respectively). We use Eqs.~(\ref{49}) and (\ref{19}) and define $f_1 =
F^2 / V^2$ and $f_2 = F^3 / V^3$. For the case of $\rho_A^{(1)}(E)$ we
obtain
\ba
\kappa^{(1)}_A(2) &=& \frac{1}{18} A V^2 (1 - f_1) \ , \nonumber \\
\kappa^{(1)}_A(3) &=& - \frac{1}{135} A V^3 (1 - f_1) (1 - 2 f_1)
\ , \nonumber \\
\kappa^{(1)}_A(4) &=& \frac{1}{135} A V^4 (1 - f_1) (1 - 6 f_1 + 6
f^2_1) , \nonumber \\
\kappa^{(1)}_A(5) &=&  - \frac{4}{1701} A V^5 (1 - f_1) (1 - 14 f_1
+ 36 f^2_1 - 24 f^3_1) \ , \nonumber \\
\kappa^{(1)}_A(6) &=& \frac{31}{20412} A V^6 (1 - f_1) (1 - 30 f_1
+ 150 f^2_1 - 240 f^3_1 + 120 f^4_1) \ ,
\label{56}
\ea
and for the case of $\rho_A^{(2)}(E)$,
\ba
\kappa^{(2)}_A(2) &=& \frac{3}{80} A V^2 (1 - f_2) \ , \nonumber \\
\kappa^{(2)}_A(3) &=& - \frac{1}{160} A V^3 (1 - f_2) (1 - 2 f_2) \ ,
\nonumber \\
\kappa^{(2)}_A(4) &=& \frac{39}{8960} A V^4 (1 - f_2) (1 - 6 f_2 +
6 f^2_2) \ , \nonumber \\
\kappa^{(2)}_A(5) &=& - \frac{3}{1792} A V^5 (1 - f_2) (1 - 14 f_2
+ 36 f^2_2 - 24 f^3_2) \ , \ \nonumber \\
\kappa^{(2)}_A(6) &=& \frac{79}{86016} A V^6 (1 - f_2) (1 - 30 f_2
+ 150 f^2_2 - 240 f^3_2 + 120 f^4_2) \ .
\label{57}
\ea
The comparison of the level density calculated with the exact
cumulants up to $\kappa_A(6)$ as in Section \ref{examples} and the
asymptotic ones shows that the agreement is very good and improves
with increasing values of particle number $A$ and level number $B$. As
an example we show in Fig.~\ref{fig6} results for
$\overline{\rho}_1^{(1)}(E)$ with $V=45$ MeV, $F=37$ MeV and $A =
100$, $B = 148$. We note the slightly larger asymmetry in the
low-energy part of the spectrum calculated with the asymptotic
values. Furthermore, an energy interval with unphysical negative level
density appears around 3400 MeV in the high-energy part of the
spectrum. A similar comparison for $\overline{\rho}_1^{(2)}(E)$ with
the same parameters $V$, $F$ and particle and level numbers $A = 200$
and $B = 360$, respectively, is shown in Fig.~\ref{fig7}. In this case
the results for the exact and the asymptotic values for the cumulants
are almost indistinguishable; the slightly more pronounced asymmetry of the
asymptotic results is barely visible. 

%

\section{Density of Particle-Hole States}
\label{ph}

To calculate pre-equilibrium processes, the densities of
$p$-particle $p$-hole states are needed in addition to the total
level density worked out in the previous Sections. A $p$-particle
$p$-hole state has $p$ particles above the Fermi energy $F$ and $(A -
p)$ particles below $F$. We now show that these densities are easily
obtained by adapting the formulas obtained previously.

For the hole states there are $(A - p)$ particles distributed over the
$A$ single-particle states in the energy interval $\{0, F \}$. We
work with these $(A - p)$ particles (instead of the $p$ holes)
throughout but use the terminology of ``hole states''. The Fermi
energy $F_h$ of the $(A - p)$ particles is defined by the analogue of
Eq.~(\ref{33}) as
\be
A - p = \int_0^{F_h} {\rm d} \ve \ \overline{\rho}_1(\ve) \ .
\label{58}
\ee
The results in Eqs.~(\ref{22}) and (\ref{47}) for the low moments and
in Eqs.~(\ref{56}), (\ref{57}) for the asymptotic cumulants are
transcribed to the case of hole states by the replacements $B \to A, A
\to (A - p), V \to F, F \to F_h$. With $i = 1, 2$ this yields the full
or the asymptotic expressions for the Fourier transforms ${\cal
  F}^{(i)}_{A - p}(\tau)$ of the densities $R^{(i)}_{A - p}(E)$. These
are normalized to unity. The full densities are given by
$\rho^{(i)}_{A - p}(E) = N^{(i)}_{A - p} R^{(i)}_{A - p}(E)$, with
$N^{(i)}_{A - p} = {A \choose A - p} = S^{(i)}$. The spectrum extends
from $E^{(i)}_{1, A - p}$ to $E^{(i)}_{{S^{(i)}}, A - p}$ and has mean
energy $E^{(i)}_{0, A - p}$. These energies are defined in analogy to
Eqs.~(\ref{42}).

For the particle states there are $p$ particles distributed over the
$(B^{(i)} - A)$ single-particle states in the energy interval $\{F, V
\}$. The Fermi energy is defined by
\be
p = \int_F^{F_p} {\rm d} \ve \ \overline{\rho}_1(\ve) \ .
\label{59}
\ee
The sums $\sum_j (\tilde{\ve}_j)^k$ are conveniently determined as the
differences between the total sums given in Eqs.~(\ref{47}) and the
sums for the hole states as determined in the previous paragraph. The
transcriptions $B \to (B^{(i)} - A), A \to p$ and Eqs.~(\ref{22}) yield
the low moments. For the asymptotic cumulants, we also replace $V \to
(V - F)$ and $F \to F_p$. This yields the full or the asymptotic forms
of the Fourier transforms ${\cal F}^{(i)}_p(\tau)$ of the normalized
densities $R^{(i)}_p(E)$. The full densities are given by
$\rho^{(i)}_p(E) = N^{(i)}_p R^{(i)}_p(E)$, with $N^{(i)}_p = {B^{(i)}
  - A \choose p} = T^{(i)}$. The spectrum extends from $E^{(i)}_{1,
  p}$ to $E^{(i)}_{{T^{(i)}}, p}$ and has mean energy $E^{(i)}_{0,
  p}$. These energies are defined in analogy to Eqs.~(\ref{42}).

The level density $\rho_{A - p, p}(E) = N^{(i)}_{A - p} N^{(i)}_p R_{A
  - p, p}(E)$ for the $p$-particle $p$-hole states is the
convolution of $\rho_{A - p}(E)$ and of $\rho_p(E)$. We replace all 
densities $\rho$ by the normalized densities $R$ and
Fourier transform the result. Then the Fourier transform ${\cal
  F}^{(i)}_{A - p, p}(E)$ of $R_{A - p, p}(E)$ is given by
\ba
{\cal F}^{(i)}_{A - p, p}(\tau) &=& \int_{- \infty}^{+ \infty} {\rm d} E \
\exp \{ i (E - E_{0, p} - E_{0, A - p}) \tau \} \int_{- \infty}^{+ \infty}
{\rm d} E_1 \int_{- \infty}^{+ \infty} {\rm d} E_2 \nonumber \\
&& \qquad \qquad \times \delta(E - E_1 - E_2) \rho^{(i)}_{A - p}(E_1)
\rho^{(i)}_p(E_2) \nonumber \\
&=& {\cal F}^{(i)}_{A - p}(\tau) {\cal F}^{(i)}_p(\tau) \ .
\label{60}
\ea
The spectrum extends from $E^{(i)}_{1, A - p} + E^{(i)}_{1, p}$ to
$E^{(i)}_{{S^{(i)}}, A - p} +E^{(i)}_{{T^{(i)}}, p}$. In this interval the
expression~(\ref{60}) can be used straightforwardly for an expansion
in orthogonal polynomials. It is gratifying to see that the
calculation of the particle-hole density requires only about the same
effort as the calculation of the total density.

\section{Density of Accessible States and of Accessible 
Particle-Hole States \label{rhoacces}}

The density $\rho_{\rm acc}(E)$ of accessible states is an important
concept in the theory of pre-equilibrium reactions. It is used to
determine the rate for transitions induced by an external agent (an
impinging proton, or a laser photon, for instance). Here we define
$\rho_{\rm acc}(E)$ for processes where an external agent (a laser
pulse, for example) excites an individual nucleon from a many-body
state at energy $E$ to another such state at energy $E + \Delta E$. We
assume that at both energies the nucleus is in thermal equilibrium and
calculate $\rho_{\rm acc}(E)$ in the framework of the Fermi-gas model
of Eqs.~(\ref{37}) and (\ref{38}). At high excitation energies this
model is much simpler to use than the usual counting procedure for
$\rho_{\rm acc}(E)$.

\begin{figure}[ht]
\vspace{5 mm}
\includegraphics[width=0.5\linewidth]{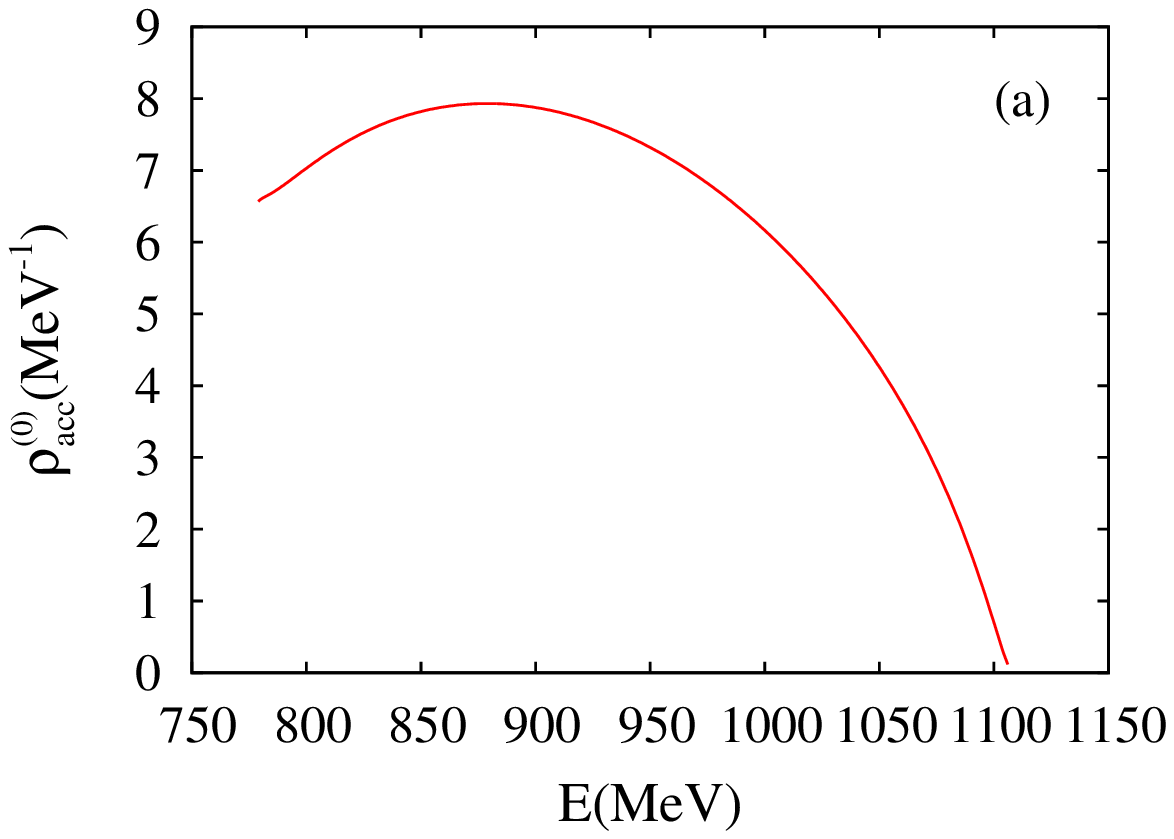}
\includegraphics[width=0.5\linewidth]{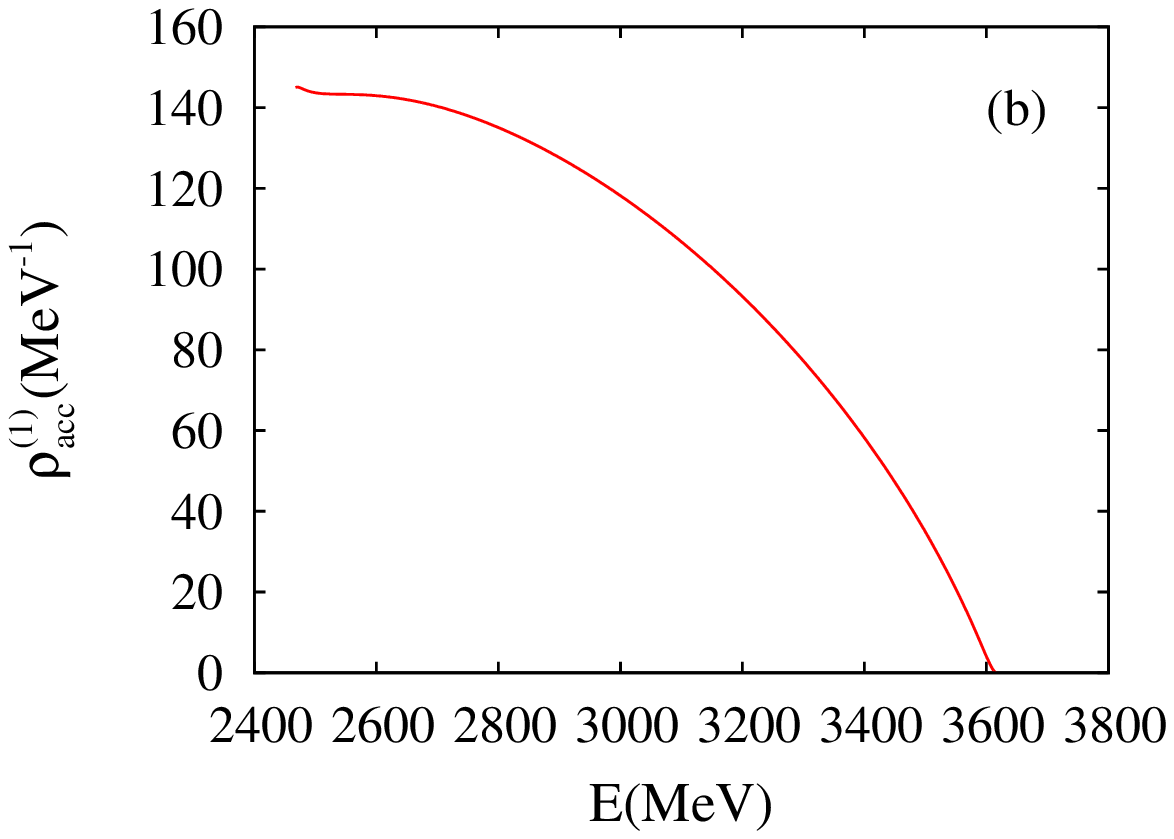}
\includegraphics[width=0.5\linewidth]{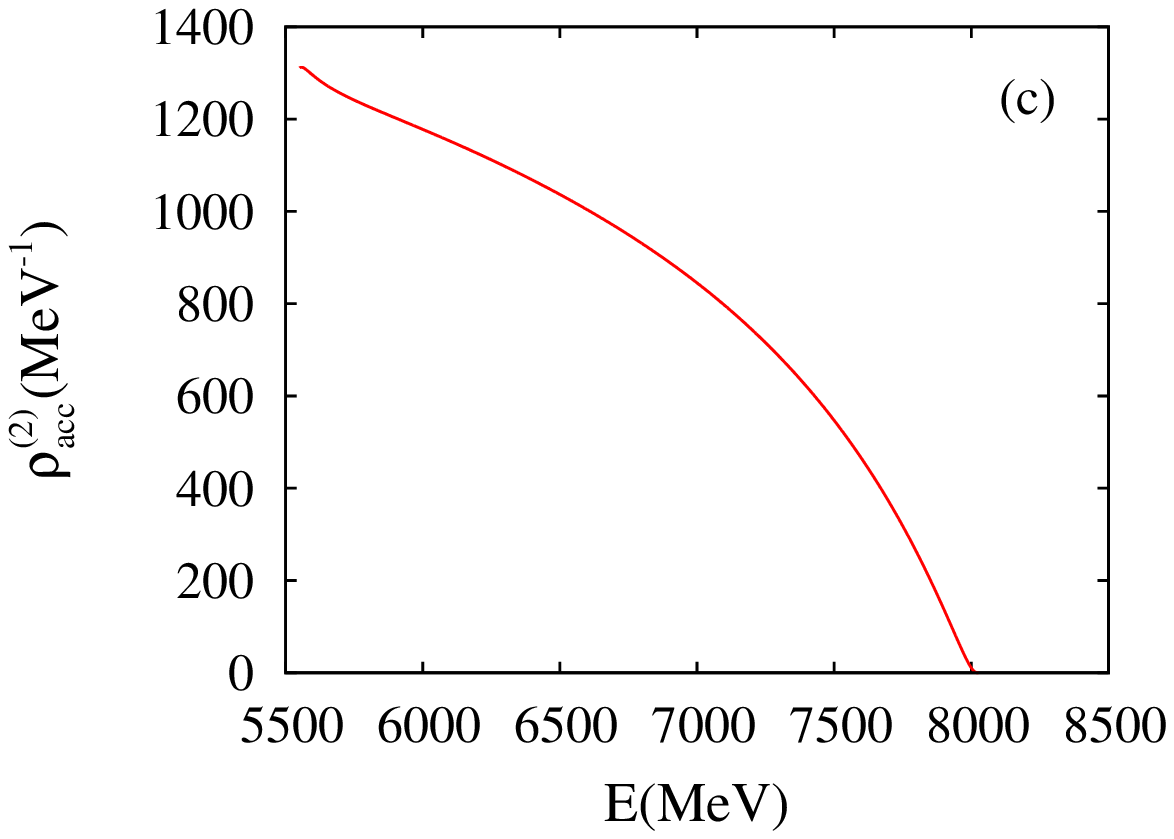}
\vspace{3 mm}
\caption{Density of accessible states $\rho_{\rm acc}(E,\Delta)$ for
  $\Delta = 5$ MeV and single-particle level densities (a)
  $\overline{\rho}_1^{(0)}$, (b) $\overline{\rho}_1^{(1)}$ and (c)
  $\overline{\rho}_1^{(2)}$ for the same parameters as used in
  Section~\ref{examples}.}
\label{fig8} 
\end{figure}

With $n_{A, E}(\ve)$ the probability of finding a single-particle
state with energy $\ve$ occupied when the nucleus has total energy $E$
and $(1 - n_{A, E + \Delta}(\ve + \Delta))$ the probability of finding
a single-particle state with energy $\ve + \Delta$ empty when the
nucleus has total energy $E + \Delta$, the number ${\cal N}_{\rm
  acc}(E, \Delta)$ of accessible states is given by the product of
both probabilities, integrated over all $\ve$ obeying $0 \leq \ve \leq
V - \Delta$,
\be
{\cal N}_{\rm acc}(E, \Delta) = \int_0^{V - \Delta} {\rm d} \ve \ n_{A,
E}(\ve) (1 - n_{A, E + \Delta}(\ve + \Delta)) \overline{\rho}_1(\ve) \ .
\label{61}
\ee
The density of accessible states is obtained by weighing the integrand
with the single-particle level density at energy $\ve + \Delta$,
\be
\rho_{\rm acc}(E, \Delta) = \int_0^{V - \Delta} {\rm d} \ve \ n_{A,
E}(\ve) (1 - n_{A, E + \Delta}(\ve + \Delta)) \overline{\rho}_1(\ve)
\overline{\rho}_1(\ve + \Delta) \ .
\label{62}
\ee
The range of $\Delta$ is $0 \leq \Delta \leq V$. Densities of
accessible states for the three cases of single-particle level
densities defined in Eqs.~(\ref{40}) are shown in Fig.~\ref{fig8}.
Obviously, the stronger the increase of $\overline{\rho}_1(\ve)$ with
single-particle energy $\ve$, the bigger the density of accessible
states. Conversely, increasing the energy $E$ of the many--body system
makes $n_{A, E}(\ve)$ (which is a step-like function of $\ve$ at low
energies) an ever more smooth function of $\ve$. At infinite
temperature, $n_{A, E}(\ve)$ is a constant independent of $\ve$. These
facts cause $\rho_{\rm acc}(E, \Delta)$ to decrease strongly with
increasing $E$.

In pre-equilibrium models, the rates for transitions at fixed energy
$E$ between states carrying different particle-hole numbers determine
the rate of equilibration of the compound nucleus. The transition
rates are proportional to the density of accessible states. We
generalize the Fermi-gas model of Eqs.~(\ref{37}) and (\ref{38}) to
the case where particle number $p$ and hole number $p$ are fixed, see
Section~\ref{ph}. We then deal with two gases in thermal equilibrium,
\be
n_{A - p, E}(\ve) = \frac{\Theta(F - \ve)}{1 + \exp \{ \beta
\ve + \alpha_{A - p} \}} \ , \ n_{p, E}(\ve) = \frac{\Theta(\ve - F)}{1
+ \exp \{ \beta \ve + \alpha_p \}} \ .
\label{63}
\ee
Here $\Theta$ is the Heaviside function. The constants $\beta,
\alpha_p$ and $\alpha_{A - p}$ are determined by the constraints
\ba
A - p &=& \int_0^F {\rm d} \ve \ n_{A - p, E}(\ve)
\overline{\rho}_1(\ve) \ , \nonumber \\
p &=& \int_F^V {\rm d} \ve \ n_{p, E}(\ve) \overline{\rho}_1(\ve) \ ,
\nonumber \\
E &=& \int_0^V {\rm d} \ve \ \ve (n_{p, E}(\ve) + n_{A - p}(\ve))
\overline{\rho}_1(\ve) \ .
\label{64}
\ea
The process leading from a $p$-particle $p$-hole state to a $(p +
1)$-particle $(p + 1)$-hole state at the same energy $E$ consists in
lifting a particle from a single-particle state at energy $\ve_1$
below the Fermi energy $F$ to a state at energy $(\ve_1 + \Delta)$ above
$F$. The required energy $\Delta$ is either taken from a particle at
energy $\ve_2 > F$ that is moved to a state at energy $\ve_2 - \Delta
> F$, or from a particle at energy $\ve_3 < F$ that is moved to a
state at energy $(\ve_3 - \Delta)$. For clarity we define
\be
\overline{\rho}_{1, <}(\ve) = \Theta(F - \ve) \overline{\rho}_1(\ve) \
, \ \overline{\rho}_{1, >}(\ve) = \Theta(\ve - F) \overline{\rho}_1(\ve)
\ .
\label{65}
\ee
We recall that $\overline{\rho}_1(\ve) = 0$ for $\ve < 0$ and for $\ve
> V$. According to Eq.~(\ref{62}), the densities of accessible states
for the three processes are
\ba
\rho_{{\rm acc}, 1}(E, \Delta) &=& \int_{- \infty}^\infty {\rm d} \ve_1
\ n_{A - p, E}(\ve_1) (1 - n_{p, E}(\ve_1 + \Delta)) \nonumber \\
&& \qquad \times \overline{\rho}_{1, <}(\ve_1) \overline{\rho}_{1, >}
(\ve_1 + \Delta) \ , \nonumber \\
\rho_{{\rm acc}, 2}(E, \Delta) &=& \int_{- \infty}^\infty {\rm d} \ve_2
\ n_{p, E}(\ve_2) (1 - n_{p, E}(\ve_2 - \Delta))
\overline{\rho}_{1, >}(\ve_2) \overline{\rho}_{1, >}(\ve_2 - \Delta) \ ,
\nonumber \\
\rho_{{\rm acc}, 3}(E, \Delta) &=& \int_{- \infty}^\infty {\rm d} \ve_3
\ n_{p, E}(\ve_3) (1 - n_{p, E}(\ve_3 - \Delta))
\overline{\rho}_{1, <}(\ve_3) \overline{\rho}_{1, <}(\ve_3 - \Delta) \ .
\nonumber \\
\label{66}
\ea
Because of the definitions~(\ref{65}), we need not restrict the
ranges of integration in Eqs.~(\ref{66}) or the range of $\Delta$. The
density of accessible $(p + 1)$-particle $(p + 1)$-hole states is
given by
\ba
\rho_{\rm acc}(E) &=& \frac{1}{2} \int_{- \infty}^\infty \ {\rm d} \Delta \
\rho_{{\rm acc}, 1}(E, \Delta) \rho_{{\rm acc}, 2}(E, \Delta) \nonumber \\
&& \qquad + \frac{1}{2} \int_{- \infty}^\infty {\rm d} \Delta \
\rho_{{\rm acc}, 1}(E, \Delta) \rho_{{\rm acc}, 3}(E, \Delta) \ .
\label{67}
\ea
The substitution $\Delta \to (\ve_2 - \ve_1 - \Delta')$ shows that the
two indistinguishable processes $\ve_1 \to (\ve_1 + \Delta)$, $\ve_2 \to
(\ve_2 - \Delta)$ and $\ve_1 \to (\ve_2 - \Delta)$, $\ve_2 \to (\ve_1 +
\Delta)$ are both contained in the first integral. To avoid double
counting we introduce the factor $1 / 2$. The same argument applies to
the processes $\ve_1 \to (\ve_1 + \Delta)$, $\ve_3 \to (\ve_3 - \Delta)$
and $\ve_3 \to (\ve_1 + \Delta)$, $\ve_1 \to (\ve_3 - \Delta)$, hence the
second factor $1 / 2$.

\section{Summary and Conclusions}

The level density is a basic property of atomic nuclei. It has been
the object of theoretical and experimental investigations since the
beginnings of nuclear physics. In the present work we have focused
attention on the domain of high excitation energies (several $100$ MeV
above yrast) and large particle numbers ($A \geq 100$). For the
theoretical description of heavy-ion reactions at several MeV per
nucleon and of reactions induced by coherent laser beams with several
MeV per photon, one needs to know the level density $\rho_A(E)$ in
that domain where (measured in units of the mean single-particle
density) $\rho_A(E)$ easily attains values of $10^{40}$ or $10^{50}$.

Starting from a set of single-particle energies given either
empirically or in terms of a mean-field approximation and generalizing
the approach developed in Ref.~\cite{Pal13}, we have have written
$\rho_A(E)$ as the sum of all ways of distributing $A$ spinless
Fermions over the available single-particle states. We have derived an
exact closed-form expression for the Fourier transform (with respect
to energy $E$) and Laplace transform (with respect to particle number
$A$) of $\rho_A(E)$. That expression yields exact values for the
lowest moments and for the lowest cumulants of $\rho_A(E)$. These
depend on binomial coefficients and on the moments of the
single-particle energies. They were used to construct approximate
expressions for the Fourier transform of $\rho_A(E)$ and, from there,
approximate expressions for the coefficients of an expansion of
$\rho_A(E)$ in terms of orthogonal polynomials. As an alternative to
using a fixed set of single-particle energies we have also considered
a smooth form of the single-particle level density
$\overline{\rho}_1(\ve)$. We have demonstrated that the approach
converges: For $A \gg 1$ and realistic forms of
$\overline{\rho}_1(\ve)$, the cumulants $\kappa_A(k)$ quickly decrease
with increasing $k$. We have shown that the huge values of the level
density attained at the center of the spectrum make it appear very
unlikely that a uniform approximation to $\rho_A(E)$ (valid throughout
the spectrum) will ever be practicable. In that sense, our approach
complements the standard approaches to calculating $\rho_A(E)$ that
focus on low excitation energies and small particle numbers. Being
entirely analytical, the present approach provides direct insight into
the overall dependence of $\rho_A(E)$ on energy. Moreover, it is easy
to implement.

For the constant-spacing model we have tested our approach against
exact numerical results from Ref.~\cite{Pal13}, and we have compared
it with approximate results of Ref.~\cite{Pal13} where the low moments
of $\rho_A(E)$ were used as fit parameters. We find good agreement in
the center of the spectrum whereas our results give too high values
for $\rho_A(E)$ in the tails. The present approach is not confined to
the (unrealistic) constant-spacing model, and we have calculated
$\rho_A(E)$ using two more realistic energy-dependent forms of
$\overline{\rho}_1(\ve)$. (The approach can easily be used for other
forms). As in the case of the constant-spacing model, our approach
fails in the tails of $\rho_A(E)$. We have shown that as the
dependence of $\overline{\rho}_1(\ve)$ on the single-particle energy
$\ve$ increases, the maximum of $\rho_A(E)$ is significantly shifted
toward higher excitation energy, the spectrum becomes asymmetric, and
the width in energy of $\rho_A(E)$ decreases.  For large particle
numbers our approximation covers about half the spectrum around the
center and a range of values of $\rho_A(E)$ covering about 20 orders
of magnitude.

Using the same technique we have also determined particle-hole
densities in the domain of large excitation energies and particle
numbers. The calculation of these quantities is as straightforward as
that of $\rho_A(E)$. It goes without saying that instead of
considering Fermions, we can use our approach also to determine level
densities in cases where neutrons and protons are considered separate
entities. For the calculation of the density $\rho_{\rm acc}$ of
accessible states we have used an equilibrated Fermi-gas model and
several forms of $\overline{\rho}_1(\ve)$. Our results are intuitively
obvious: $\rho_{\rm acc}$ increases strongly with increasing energy
dependence of $\overline{\rho}_1(\ve)$, and it decreases strongly with
increasing excitation energy.

The present approach is limited to non-interacting Fermions. The
nucleon-nucleon interaction can be partially taken into account in
terms of a mean-field approach. Even the results of a
temperature-dependent Hartree-Fock approximation can be accommodated by
readjusting the cumulants for each value of the temperature. Thus, we 
expect the method will be useful for a broad range of applications.

\appendix

\section{Moments of the Density \label{momentsApp}}

Combining Eqs.~(\ref{21}), (\ref{20}), and (\ref{15}) we
obtain~\cite{Pal13} for $k = 0, 1, \ldots, 6$
\ba
m_A(0) &=& {B \choose A} \ , \nonumber \\
m_A(1) &=& 0 \ , \nonumber \\
m_A(2) &=& {B - 2 \choose A - 1} \ \sum_{j = 1}^B (\tilde{\ve}_j)^2
\ , \nonumber \\
m_A(3) &=& \bigg\{ {B - 3 \choose A - 1 } - { B - 3 \choose A - 2 }
\bigg\} \ \sum_{j = 1}^B (\tilde{\ve}_j)^3 \ , \nonumber \\ 
m_A(4) &=& \bigg\{ {B - 4 \choose A - 1} - 4 {B - 4 \choose A - 2}
+ {B - 4 \choose A - 3} \bigg\} \ \sum_{j = 1}^B (\tilde{\ve}_j)^4 
\nonumber \\
&& + 3 {B - 4 \choose A - 2} \ \bigg( \sum_{j = 1}^B
(\tilde{\ve}_j)^2 \bigg)^2 \ , \nonumber \\
m_A(5) &=& \bigg\{ {B - 5 \choose A - 1} - 11 {B - 5 \choose A - 2}
+ 11 {B - 5 \choose A - 3} - {B - 5 \choose A - 4} \bigg\} \
\sum_{j = 1}^B (\tilde{\ve}_j)^5 \nonumber \\
&& + 10 \bigg\{ {B - 5 \choose A - 2} - {B - 5 \choose A - 3} \bigg\}
\ \sum_{j = 1}^B (\tilde{\ve}_j)^2 \sum_{j = 1}^B (\tilde{\ve}_j)^3
\ , \nonumber \\
m_A(6) &=& \bigg\{ {B - 6 \choose A - 1} - 26 {B - 6 \choose A - 2}
+ 66 { B - 6 \choose A - 3} - 26 { B - 6 \choose A - 4} \nonumber \\
&& \qquad \qquad + {B - 6 \choose A - 5} \bigg\} \ \sum_{j = 1}^B
(\tilde{\ve}_j)^6 \nonumber \\
&& + 15 \bigg\{ {B - 6 \choose A - 2} - 4 {B - 6 \choose A - 3} +
{B - 6 \choose A - 4} \bigg\} \  \sum_{j = 1}^B (\tilde{\ve}_j)^2
 \sum_{j = 1}^B (\tilde{\ve}_j)^4 \nonumber\\
&& + 10 \bigg\{ {B - 6 \choose A - 2} - 2 {B - 6 \choose A - 3} + {B
- 6 \choose A - 4} \bigg\} \ \bigg( \sum_{j = 1}^B (\tilde{\ve}_j)^3
\bigg)^2 \nonumber \\
&& + 15 {B - 6 \choose A - 3} \ \bigg( \sum_{j = 1}^B (\tilde{\ve}_j)^2
\bigg)^3 \ .
\nonumber
\ea
For the single-particle model with constant level density we extend
the calculation to $k = 8$. In that case we have $m_A(7) = 0$ and
\ba
m_A(8) &=&  \ \sum_{j = 1}^B (\tilde{\ve}_j)^8 \ \bigg\{ {B - 8
\choose A - 1} - 120 {B - 8 \choose A - 2} + 1191 {B - 8 \choose
A - 3} \nonumber \\
&& - 2416 {B - 8 \choose A - 4} + 1191 {B - 8 \choose A - 5}
- 120 {B - 8 \choose A - 6} \nonumber \\
&& \qquad + {B - 8 \choose A - 7} \bigg\} \nonumber \\
&& + 28 \  \sum_{j = 1}^B (\tilde{\ve}_j)^2 \ \sum_{j = 1}^B
(\tilde{\ve}_j)^6 \ \bigg\{ {B - 8 \choose A - 2} - 26 {B - 8
\choose A - 3} + 66 {B - 8 \choose A - 4} \nonumber \\
&& \qquad - 26 {B - 8 \choose A - 5} + {B - 8 \choose A - 6}
\bigg\} \nonumber \\
&& + 35 \ \bigg( \sum_{j = 1}^B (\tilde{\ve}_j)^4
\bigg)^2 \ \bigg\{ {B - 8 \choose A - 2} -
8 {B - 8 \choose A - 3} + 18 {B - 8 \choose A - 4} \nonumber \\
&& \qquad - 8 {B - 8 \choose A - 5} + {B - 8 \choose A - 6}
\bigg\} \nonumber \\
&& +210 \ \bigg( \sum_{j = 1}^B (\tilde{\ve}_j)^2
\bigg)^2 \ \sum_{j = 1}^B (\tilde{\ve}_j)^4 \ \bigg\{ {B - 8
\choose A - 3} - 4 {B - 8 \choose A - 4} \nonumber \\
&& \qquad + {B - 8 \choose A - 5} \bigg\} \nonumber \\
&& + 105\ \bigg( \sum_{j = 1}^B (\tilde{\ve}_j)^2
\bigg)^4 \ {B - 8 \choose A - 4} \ .
\label{22}
\ea
%

\section{In Search of a Uniform Approximation to $\rho_A(E)$ \label{searchApp}}

To justify our claim that different approximations are needed in
different parts of the spectrum, we use as a model the smooth
equivalent of the constant-spacing model, i.e, a smooth
single-particle level density $\overline{\rho}_1(E) = 1 / d$ for $- V
/ 2 \leq E \leq V / 2$ and $\overline{\rho}_1(E) = 0$ otherwise. It is
clear from the outset and shown presently that we cannot hope for
exact agreement with results obtained for a level density of the form
$\rho_1(E) = \sum_j \delta(E - d j)$. That is irrelevant, however, for
our claim. For the many-body level density $\rho_A(E)$ we use the
ansatz
\ba
\rho_A(E) &=& \frac{1}{d^A} \frac{1}{A!} \int_{- V / 2}^{V / 2} {\rm d}
E_1 \int_{- (V - d) / 2 }^ {(V - d) / 2} {\rm d} E_2 \times \ldots \times
\int_{- (V - (A - 1) d ) / 2}^{(V - (A - 1) d ) / 2} {\rm d} E_A \nonumber \\
&& \times \delta(E_1 + E_2 + \ldots E_A - E) \ .
\label{dens}
\ea
The factor $1 / A!$ takes care of the exclusion principle. The limits
of integration are chosen in such a way as to be consistent with both
the domain of definition of $\overline{\rho}_1(E)$ and the fact that
the level density vanishes for $V \leq d A$. As we shall see, this
choice also guarantees the correct normalization of $\rho_A(E)$. The
price we have to pay is that the range of $\rho_A(E)$ is not
reproduced correctly.  For the constant-spacing model that range
(after a suitable shift of energy) is given~\cite{Pal13} by $- (1 / 2)
V A + (1 / 2) d A^2 \leq E \leq (1 / 2) V A - (1 / 2) d A^2$ while
Eq.~(\ref{dens}) implies $ - (1 / 2) V A + (1 / 4) d A (A - 1) \leq E
\leq (1 / 2) V A - (1 / 4) d A (A - 1)$. The range is bigger in the
present case than in the constant-spacing model defined by a sum of
delta functions, $\rho_1(E) = \sum_j \delta(E - d j)$.

We introduce the dimensionless quantities $B = V / d$, $\epsilon = E /
d$, $\epsilon_j = E_j / d$, $j = 1, \ldots, A$ and have
\ba
d \rho_A(\epsilon) &=& \frac{1}{A!} \int_{- B / 2}^{B / 2} {\rm d}
\epsilon_1 \int_{- (B - 1) / 2}^ {(B - 1) / 2} {\rm d} \epsilon_2 \times
\ldots \times \int_{- (B - A + 1) / 2}^{(B - A + 1) / 2} {\rm d} \epsilon_A
\nonumber \\ &&  \times \delta(\epsilon_1 + \epsilon_2 + \ldots
\epsilon_A - \epsilon) \ .
\ea
We write the delta function as a Fourier integral over $\tau$ and
obtain
\ba
d \rho_A(\epsilon) &=& \frac{1}{2 \pi} \frac{1}{A!} \int_{- \infty}^{+
\infty} {\rm d} \tau \ \exp \{ - i \epsilon \tau \} \int_{- B / 2}^{B / 2}
{\rm d} \epsilon_1 \exp \{ i \tau \epsilon_1 \} \nonumber \\
&& \times \int_{- (B - 1) / 2}^{(B - 1) / 2} {\rm d} \epsilon_2 \exp \{ i
\tau \epsilon_2 \} \times \ldots \times \int_{- (B - A + 1) / 2}^{(B - A + 1)
/ 2} {\rm d} \epsilon_A \exp \{ i \tau \epsilon_A \} \nonumber \\
&=& \frac{1}{2 \pi} \frac{1}{A!} \int_{- \infty}^{+ \infty} {\rm d} \tau
\ \exp \{ - i \epsilon \tau \} \prod_{j = 0}^{A - 1}\frac{2 \sin [(B - j)
\tau] / 2 }{\tau} \nonumber \\
&=& \frac{1}{2 \pi} { B \choose A } \int_{- \infty}^{+ \infty} {\rm d}
\tau \ \exp \{ - i \epsilon \tau \} \prod_{j = 0}^{A - 1} \frac{\sin [(B
- j) \tau / 2] }{(B - j) \tau / 2} \ .
\ea
The last equality shows that $d \rho_A(\epsilon)$ is correctly
normalized. Thus, we identify
\ba
{\cal F}_A(\tau) &=& \prod_{j = 0}^{A - 1} \frac{\sin [(B - j) \tau / 2]
}{(B - j) \tau / 2} \nonumber \\
&=& \exp \bigg\{ \sum_{j = 0}^{A - 1} \ln \frac{\sin [B ( 1 - j / B ) \tau /
2] }{B (1 - j / B) \tau / 2} \bigg\}
\label{FT}
\ea
as the Fourier transform of the normalized function $R_A(\epsilon)$
with the range
\be
- (1 / 2) B A + (1 / 4) A (A - 1) \leq \epsilon \leq (1 / 2) B A
- (1 / 4) A (A - 1) \ .
\ee
The length $L$ of this interval is
\be
L = B A - (1 / 2) A (A - 1) \ ,
\ee
and the coefficients of the orthogonal polynomials have the values
\ba
r_A(n) &=& \sqrt{\frac{2}{L}} {\cal F}_A(\pi n / L) \ {\rm for} \ n \
{\rm positive \ and \ odd}, \nonumber \\
r_A(n) &=& 0 \ {\rm otherwise}.
\label{coef}
\ea

The Fourier transform ${\cal F}_A(\tau)$ is symmetric about $\tau = 0$
and attains its maximum value there. As $\tau$ increases from zero,
the first zeroes of ${\cal F}_A(\tau)$ are at
\be
\tau = \frac{2 \pi}{B}, \tau = \frac{2 \pi}{B - 1}, \tau = \frac{2
\pi}{B - 2}, \ldots, \tau = \frac{2 \pi}{B - A} \ . 
\ee
The next sequences of zeroes occur at
\be
\tau = \frac{2 k \pi}{B}, \tau = \frac{2 k \pi}{B - 1}, \tau =
\frac{2 k \pi}{B - 2}, \ldots, \tau = \frac{2 k \pi}{B - A} \ ,
\ {\rm where} \ k = 2, 3, \ldots. 
\ee
In the intervals separating these sequences, i.e., for $(2 k \pi) / (B
- A) \leq \tau \leq (2 (k + 1) \pi) / B $ we have 
\be
|{\cal F}_A(\tau)| \leq [2 / ((2 k + 1) \pi)]^A
\label{ineq}
\ee
while the arguments of ${\cal F}_A(\tau)$ needed for the calculation
of the coefficients $r_A(n)$ have the values $n \pi / L$ with $n = 1,
2, \ldots$. We display the consequences for several choices of $B$ and
$A$. For $B = 51$ and $A = 3$ the total number of states is $\approx
21 000$. To obtain a uniform approximation to $\rho_A(E)$ we must take
into account all coefficients $r_A(n)$ that are larger than $10^{-
  5}$.  According to Eq.~(\ref{ineq}) the function ${\cal F}_A(\tau)$
is smaller than that value for $k = 10$ or $\tau = 20 \pi / B$ or
$n_{\rm max} = 20 L / B \approx 60$. That is the maximum value of $n$
needed for an accurate calculation of $\rho_A(E)$ throughout the
entire spectrum. The same arguments applied to other choices of $B$
and $A$ yield for $n_{\rm max}$ the following values.
\ba
\begin{matrix} B  & A   & k_{\rm max} & n_{\rm max}  \cr
               51 & 3   & 10        & 60        \cr
               51 & 25  & 1         & 40        \cr
              100 & 30  & 1         & 50        \cr
              100 & 50  & 1         & 75        \cr
              200 & 100 & 1         & 150       \cr
\end{matrix}
\label{mat}
\ea
The number of orthogonal polynomials is perfectly manageable. The open
question is whether for $B \gg 1, A \gg 1$ it is possible to evaluate
Eqs.~(\ref{coef}) sufficiently accurately. For $B = 200, A = 100$ we
have from Stirling's formula ${B \choose A} \approx \exp \{ 200 \ln 2
\} \approx 10^{60}$. To correctly reproduce $\rho_A(\epsilon)$ also in
the tails, i.e., for $\epsilon$ near $\pm [(1 / 2) B A - (1 / 4) A (A
  - 1)]$, we need a numerical accuracy of one part in $10^{60}$. That
may be attainable but would be highly impracticable. Rewriting the
second of Eqs.~(\ref{FT}) as an integral over the continuous variable
$x = j / B$ and carrying out the integrations over $x$ yields a
closed-form expression for ${\cal F}_A(\tau)$ but does not remove the
difficulty.

\section{Moments of Single-Particle Energies \label{singpartApp}}

According to Eq.~(\ref{39a}), the moments of the single-particle
energies with $l = 0, 1, \ldots$ are
\ba
\sum_j (\ve^{(0)}_j)^l &=& \frac{1}{l + 1} A \frac{V^{l + 1}}{F} \ ,
\nonumber \\
\sum_j (\ve^{(1)}_j)^l &=& \frac{2}{l + 2} A \frac{V^{l + 2}}{F^2} \ ,
\nonumber \\
\ \sum_j (\ve^{(2)}_j)^l &=&  \frac{3}{l + 3} A \frac{V^{l + 3}}{F^3}
\ .
\label{44}
\ea
Eq.~(\ref{12}) implies $\Delta^{(0)} = \frac{1}{2} V$, $\Delta^{(1)} =
\frac{2}{3} V$, $\Delta^{(2)} = \frac{3}{4} V$, and Eq.~(\ref{11})
gives
\ba
\sum_j (\tilde{\ve}^{(0)}_j)^l &=& A \frac{V^{l + 1}}{F} \sum_{m
= 0}^l (-)^{l - m} { l \choose m} \bigg({1 \over 2}\bigg)^{l - m} {1
\over m + 1} \ , \nonumber \\
\sum_j (\tilde{\ve}^{(1)}_j)^l &=& A \frac{V^{l + 2}}{F^2} \sum_{m
= 0}^l (-)^{l - m} { l \choose m} \bigg({2 \over 3}\bigg)^{l - m} {2
\over m + 2} \ , \nonumber \\
\sum_j (\tilde{\ve}^{(2)}_j)^l &=&  A \frac{V^{l + 3}}{F^3} \sum_{m
= 0}^l (-)^{l - m} { l \choose m} \bigg({3 \over 4}\bigg)^{l - m} {3
\over m + 3} \ .
\label{45}
\ea
For $k = 0, 1, 2$ we write $1 / (m + k + 1) = \int_0^1 {\rm d} x
\ x^{m + k}$. Then the sums in Eqs.~(\ref{45}) can be carried
out, such that
\ba
\sum_{m = 0}^l (-)^{l - m} { l \choose m} \bigg({1 \over 2}\bigg)^{l
- m} {1 \over m + 1} &=& \bigg[ \frac{1}{l + 1} y^{l + 1}
\bigg]_{y = - 1/2}^{y = 1/2} \ ,
\nonumber \\
\sum_{m = 0}^l (-)^{l - m} { l \choose m} \bigg({2 \over 3}\bigg)^{l
- m} {2 \over m + 2} &=& \bigg[ \frac{2}{l + 2} y^{l + 2} + {4 \over 3
(l + 1)} y^{l + 1} \bigg]_{y = - 2/3}^{y = 1/3} \ ,
\nonumber \\
\sum_{m = 0}^l (-)^{l - m} { l \choose m} \bigg({3 \over 4}\bigg)^{l
- m} {3 \over m + 3} &=& \bigg[ \frac{3}{l + 3} y^{l + 3} + \frac{9}{2
(l + 2)} y^{l + 2} \nonumber \\
&& + \frac{27}{16 (l + 1)} y^{l + 1} \bigg]_{y = - 3/4}^{y = + 1/4} \ .
\label{46}
\ea
For the low moments we obtain
\ba
\sum_j (\tilde{\ve}^{(0)}_j)^2 &=& \frac{1}{12} A \frac{V^3}{F} \ , \
\sum_j (\tilde{\ve}^{(0)}_j)^3 = 0 \ , \
\sum_j (\tilde{\ve}^{(0)}_j)^4 = \frac{1}{80} A \frac{V^5}{F} \ , \
\nonumber \\
\sum_j (\tilde{\ve}^{(0)}_j)^5 &=& 0 \ , \
\sum_j (\tilde{\ve}^{(0)}_j)^6 = \frac{1}{448} A \frac{V^7}{F} \ ,
\nonumber \\
\sum_j (\tilde{\ve}^{(1)}_j)^2 &=& \frac{1}{18} A \frac{V^4}{F^2} \ , \
\sum_j (\tilde{\ve}^{(1)}_j)^3 = - \frac{1}{135} A \frac{V^5}{F^2} \ , \
\sum_j (\tilde{\ve}^{(1)}_j)^4 = \frac{1}{135} A \frac{V^6}{F^2} \ , \
\nonumber \\
\sum_j (\tilde{\ve}^{(1)}_j)^5 &=& - \frac{4}{1701} A \frac{V^7}{F^2} \ , \
\sum_j (\tilde{\ve}^{(1)}_j)^6 = \frac{31}{20412} A \frac{V^8}{F^2} \ ,
\nonumber \\
\sum_j (\tilde{\ve}^{(2)}_j)^2 &=& \frac{3}{80} A \frac{V^5}{F^3} \ , \
\sum_j (\tilde{\ve}^{(2)}_j)^3 = - \frac{1}{160} A \frac{V^6}{F^3} \ , \
\sum_j (\tilde{\ve}^{(2)}_j)^4 = \frac{39}{8960} A \frac{V^7}{F^3} \ , \
\nonumber \\
\sum_j (\tilde{\ve}^{(2)}_j)^5 &=& - \frac{3}{1792} A \frac{V^8}{F^3} \ ,
\sum_j (\tilde{\ve}^{(2)}_j)^6 = \frac{79}{86016} A \frac{V^9}{F^3} \ . 
\label{47}
\ea
%

\section{Asymptotic expansion\label{asympApp}}

We evaluate $L(\alpha, \tau)$ in Eq.~(\ref{21}) for $B, A \gg k$ where
$k$ is the summation index in Eq.~(\ref{21}). For simplicity we consider 
only terms with $k \gg 1$ although the
proof is not restricted to that case. We generate the terms
$\propto \exp \{ \alpha A \}$ in $L$ by expanding the exponential in
Eq.~(\ref{21}) in a Taylor series. The expansion generates
$\alpha$-dependent factors of the type ${B \choose A}^{-1} (1 + \exp
\{ \alpha \})^B f^{(n_1)} \times f^{(n_2)} \times \ldots$. The term of
zeroth order in the factors $f^{(n)}$ is ${B \choose A}^{-1} (1 + \exp
\{ \alpha \})^B$ and yields unity. For the term linear in $f^{(n)}$ we
use Eq.~(\ref{19}) and the fact that for $B, A \gg n$ the coefficient
multiplying $\exp \{ \alpha A \}$ in ${ B \choose A }^{-1} (1 + \exp
\{ \alpha \})^B f^n$ is $(A/B)^n$. Hence, the term proportional to
$f^{(n)}$ yields $\sum_{l = 1}^{n + 1} c^{(n)}_l (A/B)^l$. Similarly,
for $B, A \gg n_1 + n_2$ the term ${ B \choose A }^{-1} (1 + \exp \{
\alpha \})^B f^{(n_1)} f^{(n_2)}$ yields $\sum_{l_1 = 1}^{n_1 + 1}
c^{(n_1)}_{l_1} (A/B)^{l_1}$ $\times \sum_{l_2 = 1}^{n_2 + 1}
c^{(n_2)}_{l_2} (A/B)^{l_2}$. The argument extends to terms of higher
order. Therefore, the coefficient ${\cal F}_A(\tau)$ multiplying $\exp
\{ \alpha A \}$ in $L(\alpha, \tau)$ is asymptotically given by
\be
{\cal F}_A(\tau) = \exp \bigg\{ \sum_{k = 2}^\infty \frac{i^k
\tau^k}{k!} \sum_{l = 1}^k c^{(k - 1)}_l \bigg( \frac{A}{B}
\bigg)^l \sum_{j = 1}^B (\tilde{\ve}_j)^k \bigg\} \ .
\label{48}
\ee
Comparison with Eq.~(\ref{24}) shows that the cumulants $\kappa_A(k)$
are asymptotically given by
\be
\kappa_A(k) = \sum_{l = 1}^k c^{(k - 1)}_l \bigg( \frac{A}{B}
\bigg)^l \sum_{j = 1}^B (\tilde{\ve}_j)^k \ .
\label{49}
\ee
Each cumulant $\kappa_A(k)$ is proportional to $\sum_{j = 1}^B
(\tilde{\ve}_j)^k$. This is obviously a considerable simplification
compared to the full expression that would result from an expansion of
the logarithm in the first line of Eq.~(\ref{24}). We display the
origin of the simplification for the simplest case $k = 4$ where from
Eqs.~(\ref{25}) we have $\kappa_A(4) = M_A(4) - 3 M^2_A(2)$. We focus
attention on the coefficient of the term quadratic in $\sum_j
(\tilde{\ve}_j)^2$ in $\kappa_A(4)$ and use Eq.~(\ref{22}). The
coefficient is
\ba
&& 3 {B \choose A}^{-1} {B - 4 \choose A - 2} - 3 \bigg[ {B \choose
A}^{-1} {B - 2 \choose A - 1} \bigg]^2 \nonumber \\
&&  = 3 \frac{(A - 1)(B - A)}{B (B - 1)} \bigg[ \frac{A (B - A -
1)}{(B - 2)(B - 3)} - \frac{(A - 1) (B - A)}{B (B - 1)} \bigg] \ .
\label{50}
\ea
For $B, A \gg 1$ the term in square brackets becomes very small (it
is a sum of terms $\propto A^{-1}, B^{-1}$, and $(B - A)^{-1}$) in
comparison with the term $\propto \kappa^2_A(2)$ that contributes in
the same order in $\tau$. Therefore, the term quadratic in $\sum_j
(\tilde{\ve}_j)^2$ in $\kappa_A(4)$ is neglected in the asymptotic
approximation. The number of similar terms increases rapidly with the
index $k$ of the cumulants, and the neglect of such terms greatly
simplifies the cumulant expansion.

To show that the terms in the sum over $k$ in Eq.~(\ref{48}) decrease
rapidly with increasing $k$ we use that asymptotically ($l \gg 1$)
Eqs.~(\ref{46}) yield
\ba
\sum_{m = 0}^l (-)^{l - m} { l \choose m} \bigg({2 \over 3}\bigg)^{l
- m} {2 \over m + 2} &\rightarrow& (-)^l \frac{2}{l^2} \bigg(
\frac{2}{3} \bigg)^{l + 2} \ , \nonumber \\
\sum_{m = 0}^l (-)^{l - m} { l \choose m} \bigg({3 \over 4}\bigg)^{l
- m} {3 \over m + 3} &\rightarrow& (-)^l \frac{6}{l^3} \bigg(
\frac{3}{4} \bigg)^{l + 3} \ .
\label{51}
\ea
For the high moments this implies
\ba
\sum_j (\tilde{\ve}^{(1)}_j)^l &\to& (-)^l \frac{2}{l^2}
\frac{A}{F^2} \bigg( \frac{2 V}{3} \bigg)^{l + 2} \ , \nonumber \\
\sum_j (\tilde{\ve}^{(2)}_j)^l &\to& (-)^l \frac{6}{l^3}
\frac{A}{F^3} \bigg( \frac{3 V}{4} \bigg)^{l + 3} \ ,
\label{52}
\ea
and for the cumulants in Eq.~(\ref{49}) with the help of
Eqs.~(\ref{41})
\ba
\kappa^{(1)}_A(k) &\to& (-)^k \bigg[ \sum_{l = 1}^k c^{(k - 1)}_l
\bigg( \frac{F^2}{V^2} \bigg)^l \bigg] \frac{2}{k^2} \frac{A}{F^2}
\bigg( \frac{2 V}{3} \bigg)^{k + 2} \ , \nonumber \\
\kappa^{(2)}_A(k) &\to& (-)^k \bigg[ \sum_{l = 1}^k c^{(k -
1)}_l \bigg( \frac{F^3}{V^3} \bigg)^l \bigg] \frac{6}{k^3} \frac{A}
{F^3} \bigg( \frac{3 V}{4} \bigg)^{k + 3} \ .
\label{53}
\ea
In the Fourier transform ${\cal F}_A(\tau)$ we rescale the variable
\be
\tau \to \tilde{\tau}^{(1)} = \frac{2 V A^{1/2} \tau}{3} \ , \
\tau \to \tilde{\tau}^{(2)} = \frac{3 V A^{1/2} \tau }{4} \ .
\label{54}
\ee
The scaling absorbs the factor $A$ in the cumulants
$\kappa^{(1)}_A(2)$ and $\kappa^{(2)}_A(2)$. For $k \gg 1$ the
rescaled cumulants $\tilde{\kappa}_A(k)$ read
\ba
\tilde{\kappa}^{(1)}_A(k) &\to& (-)^k \bigg[ \sum_{l = 1}^k c^{(k -
1)}_l \bigg( \frac{F^2}{V^2} \bigg)^l \bigg] \frac{2}{k^2} \bigg(
\frac{2 V}{3 F} \bigg)^2 A^{- k/2} \ , \nonumber \\
\tilde{\kappa}^{(2)}_A(k) &\to& (-)^k \bigg[ \sum_{l = 1}^k c^{(k -
1)}_l \bigg( \frac{F^3}{V^3} \bigg)^l \bigg] \frac{6}{k^3} \bigg(
\frac{3 V}{4 F} \bigg)^3 A^{- k/2} \ .
\label{55}
\ea
The factors multiplying $A^{- k/2}$ in Eqs.~(\ref{55}) are of order
unity. Therefore, the rescaled cumulants fall off very rapidly with
increasing $k$ for $A \gg 1$. This justifies our asymptotic expansion
and shows that only a small number of cumulants is needed for a
reliable calculation of ${\cal F}_A(\tau)$.


\end{document}